\documentclass[twocolumn,superscriptaddress,amsmath,amssymb,floatfix,aps,pre, showpacs]{revtex4}
\usepackage{graphics,graphicx}
\usepackage{epstopdf}

\usepackage[usenames]{color}

\begin{document}

\title{Response of thermalized ribbons to pulling and bending}

\author{Andrej Ko\v{s}mrlj}
\email{andrej@princeton.edu}
\thanks{Now at {\it Princeton University, Mechanical and Aerospace Engineering, Princeton, NJ 08544}}
\affiliation{Department of Physics, Harvard University, Cambridge, MA 02138}

\author{David R. Nelson}
\email{nelson@physics.harvard.edu}
\affiliation{Department of Physics, Harvard University, Cambridge, MA 02138}
\affiliation{Department of Molecular and Cellular Biology, and School of Engineering and Applied Science, Harvard University, Cambridge, Massachusetts 02138}

\date{\today}

\begin{abstract}

Motivated by recent free-standing graphene experiments, we show how thermal fluctuations affect the mechanical properties of microscopically thin solid ribbons. A renormalization group analysis of flexural phonons reveals that elongated ribbons behave like highly anisotropic polymers, where the two dimensional nature of ribbons is reflected in non-trivial power law scalings of the persistence length and effective bending and twisting rigidities with the ribbon width. With a coarse-grained transfer matrix approach, we then show how thermalized ribbons respond to pulling and bending forces over a wide spectrum of temperatures, forces and ribbon lengths.

\end{abstract}
\maketitle

Over the last few decades the effects of thermal fluctuations acting on one dimensional polymers and two dimensional solid membranes have been studied extensively. It is well known that polymers remain straight only at short distances, while on distances larger than persistence length $\ell_p$ polymers perform a self-avoiding random walk~\cite{gennesBpolymers, doiB}. On the other hand, because of strong thermal renormalizations triggered by flexural phonons~\cite{nelson87}, arbitrarily large two dimensional membranes remain flat at low temperatures, with strongly scale-dependent enhanced bending rigidities and reduced in-plane elastic constants~\cite{nelsonB, katsnelsonB}.

A related scaling law for the membrane structure function of a solution of spectrin skeletons of red blood cells was checked in an \emph{ensemble-averaged} sense via elegant X-ray and light scattering experiments~\cite{schmidt93}.  However, recent advances in growing and isolating free-standing layers of crystalline materials such as graphene, BN, WS$_2$ or MoS$_2$~\cite{novoselov05} (not adsorbed onto a bulk substrate or stretched across supporting structure) hold great promise for exploring how flexural modes affect the mechanical properties of \emph{individual} sheet polymers that are atomically thin. Graphene also offers the opportunity to study how soft flexural phonons affect the electron transport under various conditions~\cite{mariani08, tikhonov14}, and there is a prediction of a buckling instability in hole-doped graphene~\cite{gazit09B}. Experiments carried out in a vacuum (as opposed to membranes embedded in a liquid solvent) can be extended to very low temperatures, where the quantization of in-plane and flexural phonon modes becomes important~\cite{kats14,amorim14}.

Here, inspired by recent work by Blees \emph{et al.},~\cite{blees15} we consider thermal fluctuations of microscopically thin solid ribbons of width $W$ and length $L\gg W$. We show that sufficiently long ribbons  behave like highly anisotropic one dimensional polymers, with the two-dimensional nature reflected in very large renormalizations of bending and twisting rigidities at the scale of the ribbon width $W$. It is natural to coarse-grain and construct a ribbon with $L/W\gg 1$ square membrane blocks of size $W \!\! \times \!\! W$. Below we make this idea precise, by integrating out all fluctuations on scales smaller than the ribbons' width. The work of Blees \emph{et al.}~\cite{blees15} focuses on the deflections and thermal fluctuations of free-standing graphene in the cantilever mode, and found a renormalized bending rigidity for $10\mu\textrm{m}$ wide ribbons at room temperature $\sim$6000 times larger than its microscopic value at $T=0$. \footnote{This remarkable stiffening of the bending rigidity may be due to a combination of thermal fluctuations and quenched in ripples.  See A. Ko\v{s}mrlj and D. R. Nelson, Phys. Rev. E {\bf 89}, 022126 (2014) and the review of L. Radzihovsky in Ref. [4].   Here, we focus on how thermal fluctuations alone affect the statistical mechanics of ribbons.} Although these ribbons were much shorter than the persistence length $\ell_p$, which is on the order of meters (see below), it is possible to reach the semi-flexible regime (ribbon length $L \gtrsim \ell_p$) for narrower graphene nano-ribbons. With narrower free-standing ribbons in mind, we use a coarse-grained transfer matrix approach to analyze the response of thermalized ribbons to pulling and bending for the wide spectrum of temperatures, forces and ribbon lengths.

To properly define the relevant quantities, we first discuss thermal fluctuations of large two dimensional membranes under an external edge tension $\sigma_{ij}$. The free energy cost associated with small deformations of membranes around the reference flat state is~\cite{landauB}
\begin{eqnarray}
E \!\! &=& \!\!\int \! dxdy \ \frac{1}{2}\left[\lambda u_{ii}^2 + 2 \mu u_{ij}^2 + \kappa K_{ii}^2 - 2 \kappa_G \det (K_{ij})  \right]\nonumber \\
&&-\oint \! dr\, \hat m_i \sigma_{ij} u_j,
\label{eq:free_energy_membrane}
\end{eqnarray}
where first two terms describe the cost of stretching, shearing and compressing, and the next two terms describe the cost of membrane bending. The boundary integral corresponds to the work done by external tension ($\hat m_i$ describes the unit normal vector in the X-Y plane to the membrane boundary), and summation over all indices $i,j \in \{x,y\}$ is implied. The strain tensors
$u_{ij}=(\partial_i u_j + \partial_j u_i)/2 + (\partial_i f) (\partial_j f)/2$ and
$K_{ij}=\partial_i \partial_j f$,
measure deformations from the preferred flat metric and zero curvature respectively; we kept only the lowest orders in terms of the  in-plane phonon deformations $u_i(x,y)$ and  out-of-plane deformations $f(x,y)$~\cite{landauB}.

The effects of thermal fluctuations are reflected in correlation functions obtained from functional integrals~\cite{nelsonB, katsnelsonB},
$G_{u_i u_j} ({\bf r}_2 - {\bf r}_1) = \frac{1}{Z} \int\!\! \mathcal{D}[u_i, f] \, u_i({\bf r}_2) u_j({\bf r}_1) e^{-E/k_B T}$ and 
$G_{ff} ({\bf r}_2 - {\bf r}_1)=\frac{1}{Z} \int\!\! \mathcal{D}[u_i, f] \, f({\bf r}_2) f({\bf r}_1) e^{-E/k_B T}$,
 where  $T$ is temperature, $Z=\int\!\! \mathcal{D}[u_i, f] e^{-E/k_B T}$ is the partition function and ${\bf r}=(x,y)$. In the absence of external tension ($\sigma_{ij}\equiv 0$), it is known that non-linear couplings of strain tensor $u_{ij}$ through the out-of-plane flexural phonon deformations $f(x,y)$ produce universal power law scalings of correlation functions $G({\bf q}) = \int \! (d^2{\bf r}/A) \, e^{-i {\bf q} \cdot {\bf r}} G({\bf r})$ in the long wavelength limit
$G_{u_i u_j} ({\bf q}) \sim q^{-2-\eta_u}$ and
$G_{ff} ({\bf q}) \sim q^{-4+\eta}$,
where $A$ is membrane area,  $\eta\approx 0.82$~\cite{nelson87, aronovitz88, guitter89, ledoussal92} and the exponents $\eta_u + 2 \eta=2$ are connected via Ward identities associated with the rotational symmetry~\cite{guitter89}. Thermal fluctuations become important on scales $q^{-1}$ larger than thermal length~\cite{nelson87, kantor87, aronovitz88, guitter89, ledoussal92},
\begin{equation}
\ell_\textrm{th} \sim \kappa/\sqrt{k_B T Y},
\label{eq:thermal_length}
\end{equation}
where $Y=4 \mu (\mu+\lambda)/(2 \mu +\lambda)$ is Young's modulus, and correlation functions above can be interpreted as scale dependent elastic moduli $\kappa(q),\kappa_G(q) \sim q^{-\eta}$ and $\lambda(q),\mu(q) \sim q^{+\eta_u}$~\cite{nelsonB, katsnelsonB}. Bending rigidities thus diverge for large membranes, while in-plane elastic constants become extremely small.

In order to see the role of external tension $\sigma_{ij} \ne 0$, which will help us understand pulling forces in ribbons, it is convenient to integrate out the in-plane degrees of freedom and study $E_\textrm{eff} = - k_B T \ln \left(\int \! \mathcal{D}[u_i] \, e^{-E/k_B T} \right)$, the effective free energy for out-of-plane deformations,~\cite{nelsonB}
$E_\textrm{eff}  =  \!\int \! dxdy\, \big[ (\kappa/2) \left(\nabla^2 f\right)^2 - \kappa_G \det (\partial_i \partial_j f)  +   \sigma_{ij} (\partial_i f)  (\partial_j f) +(Y/8) \big( P^T_{ij} (\partial_i f) (\partial_j f) \big)^2\big]$,
where the transverse projection operator reads $P^T_{ij}=\delta_{ij} - \partial_i \partial_j / \nabla^2$. In the effective free energy description above we see that external tension suppresses out-of-plane fluctuations in $f$, which have long range anharmonic interactions between transverse tilt deformations of the membrane normals. The effects of the anharmonic term at a given scale $\ell^*=2 \pi/ {q^*}$ can be obtained by integrating out all degrees of freedom on smaller scales. Formally this is done by splitting all fields $g({\bf r}) \in \{u_i({\bf r}), f({\bf r})\}$ into slow modes $g_<({\bf r}) = \sum_{|{\bf q}| < q^*} e^{i {\bf q} \cdot {\bf r}} g({\bf q})$ and fast modes $g_>({\bf r}) = \sum_{|{\bf q}| > q^*} e^{i {\bf q} \cdot {\bf r}} g({\bf q})$, which are then integrated out as
$E(\ell^*) = - k_B T \ln \left(\int \! \mathcal{D}[u_{i>},f_>] \, e^{-E/k_B T} \right).$
The functional integrals following from standard perturbative renormalization group calculations~\cite{aronovitz88, guitter89, amitB} lead to a free energy with the same form as in Eq.~(\ref{eq:free_energy_membrane}) except that renormalized elastic constants $\lambda_R(\ell^*), \mu_R(\ell^*),\kappa_R(\ell^*), \kappa_{GR}(\ell^*)$ become scale dependent, while the external tension $\sigma_{ij}$ remains intact.

For a small isotropic external tension $\sigma_{ij} \equiv \sigma \delta_{ij}$, or for a small uniaxial tension in the $x$-direction $\sigma_{ij} \equiv \sigma \delta_{ix} \delta_{jx}$, the tension becomes relevant on scales larger than~\cite{roldan11}
\begin{equation}
\ell_{\sigma} \sim \left(\frac{\kappa}{\sigma \ell_\textrm{th}^\eta} \right)^{1/(2-\eta)},
\label{eq:tension_length}
\end{equation}
where exponent $\eta\approx 0.82$ and thermal length scale $\ell_\textrm{th}$ [see Eq.~(\ref{eq:thermal_length})] have been defined above for membranes without external tension. As shown in the supplemental materials, external tension then produces the renormalized elastic constants
\begin{eqnarray}
\frac{\kappa_R(\ell)}{\kappa}, \frac{\kappa_{RG}(\ell)}{\kappa_G} &\sim&  \left\{ 
\begin{array}{c l}
1, & \ell < \ell_\textrm{th} \\
(\ell/ \ell_\textrm{th})^{\eta}, & \ell_\textrm{th} < \ell < \ell_{\sigma} \\
(\ell_\sigma/\ell_\textrm{th})^{\eta}, & \ell_{\sigma} < \ell
\end{array}
\right. , \nonumber \\
\frac{\lambda_R(\ell)}{\lambda}, \frac{\mu_R(\ell)}{\mu} &\sim&  \left\{ 
\begin{array}{c l}
1, & \ell < \ell_\textrm{th} \\
(\ell/\ell_\textrm{th})^{-\eta_u}, & \ell_\textrm{th} < \ell < \ell_{\sigma} \\
(\ell_\sigma/\ell_\textrm{th})^{-\eta_u} , & \ell_{\sigma} < \ell
\end{array}
\right. .
\label{eq:renormalized_elastic_constants}
\end{eqnarray}
The out of plane correlation function then becomes
$G^{-1}_{ff} ({\bf q}) = \frac{A}{k_B T} \left[ \kappa_R(2 \pi/q)\, {q}^4 + \sigma_{ij} q_i q_j \right].$
For isotropic external tensions this result agrees with Roldan \emph{et al.}~\cite{roldan11}, but the results for uniaxial external tension appear to be new. With a uniaxial tension, the long wave length $f({\bf q})$ fluctuations behave like the layer displacements of a defect-free two dimensional smectic liquid crystal~\cite{toner81}, with fluctuations along the direction $\hat{\bf x}$ of the pulling force having a reduced amplitude
$G^{-1}_{ff} \left(|{\bf q}| < \ell_\sigma^{-1}\right) \sim \frac{A}{k_B T} \left[\sigma q_x^2  + \kappa {q_y}^4  (\ell_\sigma/\ell_\textrm{th})^{\eta}  \right].$

For sufficiently large external tension
$\sigma \gtrsim k_B T Y/\kappa \equiv \sigma^*$,~\cite{roldan11}
which corresponds to $\ell_\textrm{th} \gtrsim \ell_\sigma$, thermal fluctuations become irrelevant and the renormalized elastic constants are approximately equal to the microscopic ones. Remarkably, for graphene membranes with $\kappa=1.1 \textrm{eV}$~\cite{fasolino07} and $Y=340 \textrm{N}/\textrm{m}$~\cite{lee08}, the thermal length  at room temperature is of order the lattice constant,  $\ell_\textrm{th} \sim 1 \textrm{\AA}$!~\cite{katsnelsonB, blees15} Therefore thermal fluctuations are important for all experimental situations in this case provided only that the external membrane tension is smaller than  $\sigma^* \sim 10 \textrm{N}/\textrm{m}$.

We now study ribbons of width $W$ and length $L$ that lie on average in the X-Y plane with long axis in $\hat x$ in direction and with a pulling force $F=W \sigma_{xx}$ on the ribbon end. Once we integrate out all degrees of freedom on scales smaller than $W$, the resulting strain tensors $u_{ij}$ and $K_{ij}$ depend only on the $x$ coordinate, and the renormalized elastic constants in  Eq.~(\ref{eq:renormalized_elastic_constants}) are evaluated at $\ell=W$. This results in an effectively one dimensional free energy model for the ribbon
\begin{eqnarray}
E \!\!  \! &=& \! \!\! \int_0^L \!\!\!\! dx\, \frac{W}{2}\left[\lambda_R u_{ii}^2 + 2 \mu_R u_{ij}^2 + \kappa_R K_{ii}^2 - 2 \kappa_{GR} \det (K_{ij})  \right]\nonumber \\
&&- F u_x(L).
\label{eq:free_energy_ribbon_renormalized}
\end{eqnarray}
If we then continue integrating out degrees of freedom on scales larger than $W$ in this effective one dimensional problem (see supplemental materials), the renormalized bending moduli $\kappa_R, \kappa_{GR}$ and the renormalized shear modulus $\mu_R$ remain constant. However, the in-plane elastic modulus $2\mu_R+\lambda_R$, which is related to the deformations $u_x(x,y)$ averaged over the $y$-direction, becomes smaller and smaller, a sign that the ribbon does not remain straight. In this regime the free energy description of small deformations around the flat state in Eq.~(\ref{eq:free_energy_ribbon_renormalized}) breaks down for small $F$.

\begin{figure}[t]
\includegraphics[scale=.26]{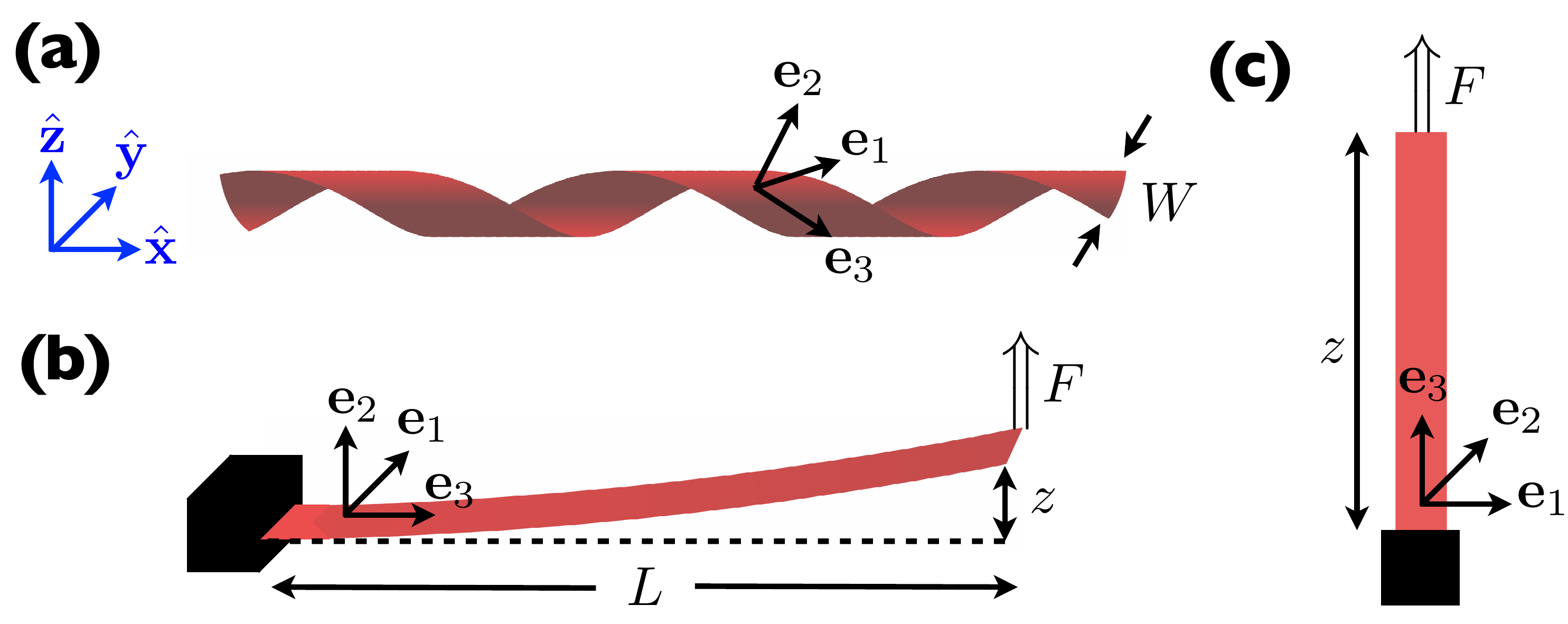}
\caption{(Color online) (a) Ribbon configurations with undeformed length $L>W$ can be described with orientations of material frame $\{{\bf e}_1, {\bf e}_2,{\bf e}_3\}$ attached to the ribbon relative to the fixed laboratory frame $\{\hat {\bf x},\hat {\bf y},\hat {\bf z}\}$. (b-c) Initial ribbon orientations for studying the response $\langle z \rangle $ to external bending and pulling forces $F$.
}
\label{fig:ribbon}
\end{figure}

For long ribbons $L \gtrsim W$ we exploit a complementary description that allows for large deformations in three dimensional space, provided that local strains remain small. We attach a material frame $\{{\bf e}_1(s), {\bf e}_2(s),{\bf e}_3(s)\}$ to the ribbon (see Fig.~\ref{fig:ribbon}), where $s\in[0,L]$ is the coordinate along the center of the ribbon backbone. The orientation of material frame relative to the fixed laboratory frame $\{\hat {\bf x},\hat {\bf y},\hat {\bf z}\}$ can be described with  Euler angles ${\bf \Theta}(s)\equiv\{\phi(s), \theta(s), \psi(s)\}$~\cite{landauBquantummechanics}.  The laboratory frame can be rotated to the local material frame with three successive three-dimensional rotations $\textrm{R}({\bf \Theta})\equiv\textrm{R}_z(-\psi) \textrm{R}_y(-\theta) \textrm{R}_z(-\phi)$~\cite{moroz98, panyukov00}, such that
$\{{\bf e}_1, {\bf e}_2,{\bf e}_3\} = \{\textrm{R}({\bf \Theta}) \hat {\bf x}, \textrm{R}({\bf \Theta})\hat {\bf y}, \textrm{R}({\bf \Theta}) \hat {\bf z}\}.$

Ribbon bending and twisting deformations are then described by the rate of rotation of the material frame along the ribbon backbone
$\frac{d{\bf e}_\alpha}{ds} = \frac{d\textrm{R}}{ds} \textrm{R}^{-1} {\bf e}_\alpha \equiv {\bf \Omega} \times {\bf e}_\alpha,$~\cite{marko94, moroz98}
with ${\bf \Omega}(s) = \Omega_\alpha {\bf e}_\alpha$. Here $\Omega_1^{-1}(s)$ and $\Omega_2^{-1}(s)$ are the radii of curvatures for bending of ribbon around axes ${\bf e}_1(s)$ and ${\bf e}_2(s)$, and $2 \pi \Omega_3^{-1}(s)$ describes the pitch for ribbon twisting. The free energy cost of a ribbon deformation is then written as~\cite{marko94, moroz98, panyukov00}
\begin{equation}
E = \int_0^L\!\!\!\! ds\, \frac{1}{2} \left[A_1 \Omega_1^2 + A_2 \Omega_2^2 + C \Omega_3^2 \right] - {\bf F} \cdot {\bf r}(L),
\label{eq:ribbon_energy}
\end{equation}
where $A_1$, $A_2$ are bending rigidities, $C$ is torsional rigidity and $F$ is the applied force on a ribbon end at ${\bf r}(L)$, which can be a bending or pulling force (see Fig.~\ref{fig:ribbon}). From comparison with the effective one dimensional model for ribbon in Eq.~(\ref{eq:free_energy_ribbon_renormalized}) we find
\begin{equation}
A_1 = W \kappa_R(W), \quad \quad C=2 W \kappa_{GR}(W),
\label{eq:renormalized_bending_twisting}
\end{equation}
where renormalized bending rigidities are defined in Eq.~(\ref{eq:renormalized_elastic_constants}). The second bending rigidity for bending around axis ${\bf e}_2(s)$, involves ribbon stretching and is much larger; in fact, $A_2$'s bare value exceeds $A_1$ and $C$ by a factor of order $Y W^2/\kappa$, the F\"oppl-von Karman number, where we expect $\kappa_G \sim \kappa$ for graphene both microscopically~\cite{landauB} and when thermal renormalizations are accounted for. This quantity can be estimated from classical zero temperature solid mechanics~\cite{landauB} as $A_2 \sim W^3 Y_R(W)$. Here, the renormalized Young's modulus $Y_R=4 \mu_R (\mu_R + \lambda_R)/(2 \mu_R + \lambda_R)$, scales in the same way as other in-plane elastic constants $\lambda_R$ and $\mu_R$ in Eq.~(\ref{eq:renormalized_elastic_constants}). For ribbons whose width is much larger than it's thickness we thus find $A_2 \gg A_1, C$ and we can set $\Omega_2 \approx 0$. In the free energy cost for ribbon deformations [Eq.~(\ref{eq:ribbon_energy})], we neglected the stretching/compressing of ribbon backbone, as is appropriate when the pulling force resisting entropic contraction is not too large~\cite{marko95}. The effective one dimensional free energy model presented above corresponds to the highly asymmetric 1d polymer~\cite{marko94, moroz98, panyukov00}, with anomalous $W$-dependent elastic parameters.~\footnote{A similar model, with $A_1 = A_2$, would describe the bending and twisting energies of hollow graphene nanotubes.  However, in this case all three elastic constants involve elastic stretching and are hence very large:  from Ref.~\cite{landauB} we have, for a cylinder of radius $R$ in terms of 2d elastic parameters, $A_1 = A_2 = \pi Y R^3$ and $C = 2\pi\mu R^3$. The $R^3$ dependencies lead to enormous persistence lengths at room temperature ($\sim$hundreds of kilometers at room temperature when $R = 1 \mu\textrm{m}$), in contrast to the behavior of thermalized ribbons, which are dominated by much softer bending deformations.
}

The free energy described by Eq.~(\ref{eq:ribbon_energy}) is complicated because successive rotations do not commute! However, the physics can be understood by mapping the statistical mechanical problem to the corresponding quantum mechanical problem~\cite{yamakawa76} as described below.

The response $\langle z \rangle $ of the ribbon to external force $F$ in the $\hat {\bf z}$ direction can be evaluated from the relation
$\langle z \rangle = k_B T (\partial \ln Z/\partial F)$,
where the partition function reads
$Z = \int \! \mathcal{D}[{\bf \Theta}(s)] e^{-E/k_B T}.$
Note that we can study both pulling and bending forces, where the only difference is the conditions on the Euler angles, i.e. in the initial orientation of ribbon (see Fig.~\ref{fig:ribbon}). If we clamp the ribbon at the origin ($s=0$) and apply force on the ribbon end ($s=L$), then for pulling the initial condition is ${\bf \Theta}_i=\{0,0,0\}$. To treat bending, we consider a ribbon initially aligned with the $\hat{\bf x}$-axis and take ${\bf \Theta}_i=\{\pi/2,\pi/2,0\}$. To evaluate the partition function $Z$, it is convenient to define the unnormalized probability distribution $\rho({\bf \Theta},s)$ of Euler angles  ${\bf \Theta}$ at a contour length $s$  along the ribbon midline as
$\rho({\bf \Theta}_f,s_f) = \int_{{\bf \Theta}(s=0)={\bf \Theta}_i}^{{\bf \Theta}(s=s_f)={\bf \Theta}_f}   \! \mathcal{D}[{\bf \Theta}(s)] e^{-E/k_B T},$
where the path integral above is restricted to $s \in [0,s_f]$ and the partition function is given by $Z = \int \!\! {d{\bf \Theta}} \rho({\bf \Theta},L)$ with the Euler-angle measure $\int \! d{\bf \Theta}\equiv\int_0^{2\pi} \!d\phi \int_0^\pi\! \sin \theta d\theta \int_0^{2 \pi} \!d\psi $. The evolution of this probability distribution along the ribbon backbone is described with differential equation
$\left(\frac{\partial}{\partial s} + \hat H \right) \rho({\bf \Theta},s) = 0,$~\cite{moroz98,eslami08}
where the Hamiltonian operator is
$\hat H = \frac{k_B T}{2} \left(\frac{\hat J_1^2}{A_1} + \frac{\hat J_2^2}{A_2} + \frac{\hat J_3^2}{C}\right) - \frac{F ({\bf e}_3 \cdot \hat {\bf z})}{k_B T}$.
Here the $\{\hat J_\alpha\}$ are angular momentum operators around axes ${\bf e}_\alpha$, which can be expressed in terms of derivatives with respect to  Euler angles~\cite{eslami08, landauBquantummechanics}. The evolution of $\rho({\bf \Theta},s)$ with $s$ maps the physics of thermalized ribbons onto the Schr\"odinger equation of the asymmetric rotating top~\cite{landauBquantummechanics} in an external gravitational field, where the ribbon backbone coordinate $s$ plays a role of imaginary time and the bending and twisting rigidities $A_1$, $A_2$ and $C$ correspond to moments of inertia. The evolution of the material frame orientation distribution can be evaluated by expanding the initial condition in eigen-distributions,  $\rho({\bf \Theta},0)=\delta({\bf \Theta} - {\bf \Theta}_i)=\sum_a C_a \rho_a ({\bf \Theta})$, where $\hat H \rho_a({\bf \Theta}) = \lambda_a \rho_a({\bf \Theta})$. In this decomposition the partition function becomes
$Z = \sum_a C_a e^{-\lambda_a L} \int \! d{\bf \Theta} \rho_a ({\bf \Theta})$
and the response $\langle z \rangle$ to an external force
can be evaluated from $Z$ as described above.

\begin{figure}[t!]
\includegraphics[scale=.30]{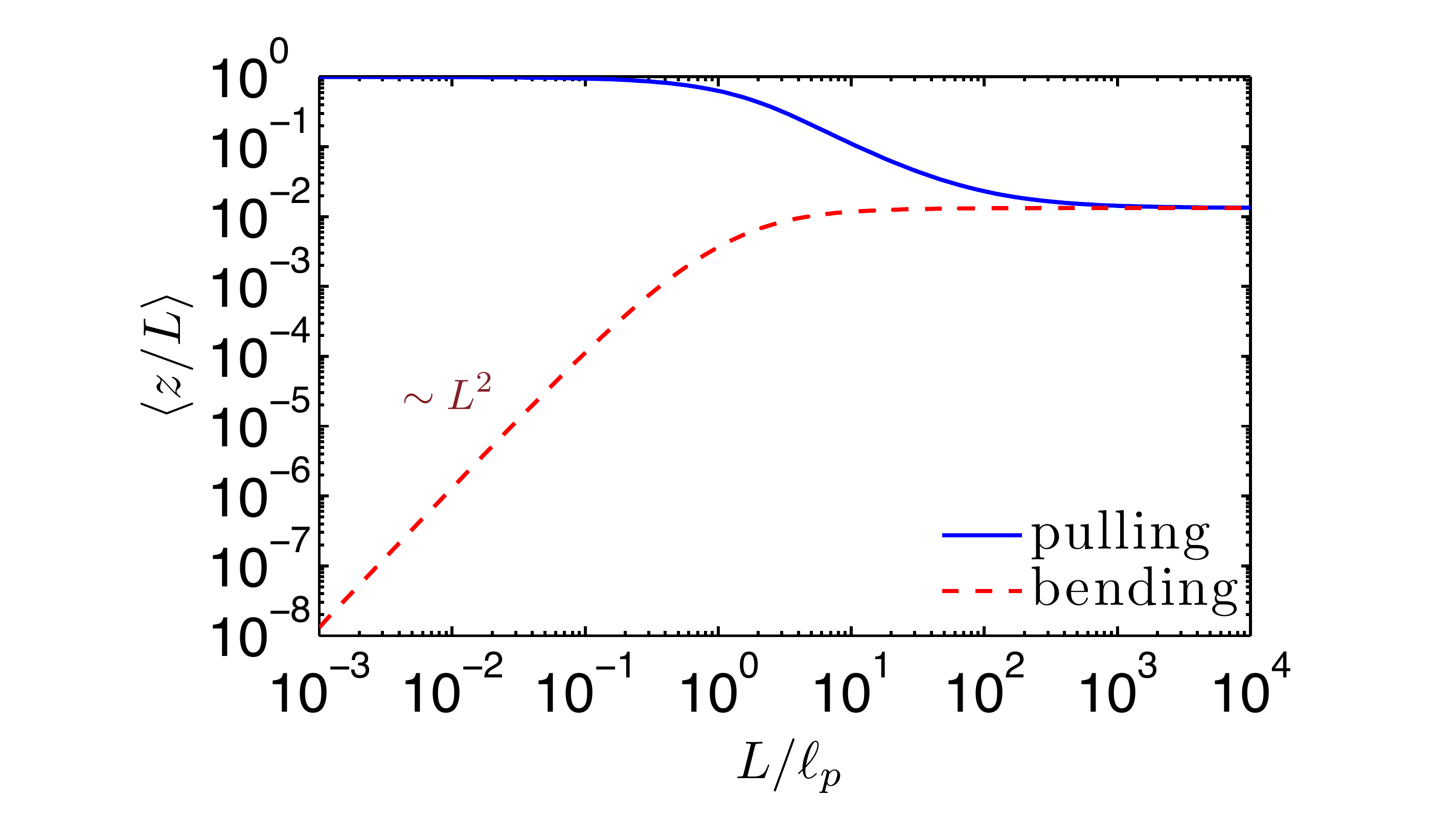}
\caption{(Color online) Pulling and bending deflections $\langle z/L \rangle$ of ribbons with bending rigidities $A_2/A_1 \rightarrow \infty$ and twisting rigidity $C/A_1=1$ in response to a fixed small external force $F A_1/(k_B T)^2 =0.01$. The slope of $+2$ for bending when $L \ll \ell_p$ agrees with expectations for stiff cantilevers with, however, a bending rigidity greatly enhanced by a factor $(W/\ell_\textrm{th})^\eta \gg 1$. The responses to pulling and bending forces agree when $L \gg \ell_p$.}
\label{fig:pulling_bending}
\end{figure}

To treat ribbons in both the semiflexible and highly crumpled regimes, we must now find all eigenvalues $\lambda_a$ and eigen-distributions $\rho_a({\bf \Theta})$. From quantum mechanics we know that this is done efficiently in the basis of Wigner D functions $D^j_{mk} ({\bf \Theta})$~\cite{landauBquantummechanics}, which have well defined quantum numbers $j,k,m$ for the total angular momentum $\hat J^2 = \hat J_1^2 + \hat J_2^2 + \hat J_3^2$, the angular momentum around the ribbon tangent $\hat J_3$ and for the angular momentum around the laboratory axis $\hat J_z$. For details see Refs.~\cite{moroz98,eslami08} and the supplemental materials.

With the help of this machinery we first studied the response of ribbons of various lengths to small external pulling and bending forces at fixed temperature (see Fig.~\ref{fig:pulling_bending}). Here, since $C$ and $A_1$ have a similar order of magnitude, we take $C=A_1$, for simplicity.
Similar to single molecule polymer physics,~\cite{bustamante03, moerner03} we find two regimes. For ribbons much shorter than a persistence length~\cite{panyukov00}
\begin{equation}
\ell_p = \frac{2}{k_B T (A_1 ^{-1} + A_2^{-1})} \approx \frac{2 W \kappa_R(W)}{k_B T}.
\end{equation}
ribbons behave like stiff ``classical rods''~\cite{landauB}, where for pulling $\langle z \rangle \approx L$  and for the bending (cantilever) mode $\langle z \rangle = F L^3/3 A_1$. Note that $A_2^{-1}$ is negligible and that thermal fluctuations on scales less than $W$ lead to a renormalized bending rigidity $A_1$ [see Eq.~(\ref{eq:renormalized_bending_twisting})], orders of magnitude larger than for rod-like polymers at room temperature, as found by the Cornell experiments~\cite{blees15}. For ribbons much longer than the persistence length ($L \gg \ell_p$), pulling and bending become equivalent. In this semi-flexible regime ribbon forgets its initial orientation after a persistence length, and for small pulling forces the response to either bending or pulling is $\langle z/L \rangle = 2 F \ell_p/(3 k_B T)$~\cite{doiB}. Eventually, at much larger ribbon lengths than those considered here, ribbon self-avoidance will become important.~\cite{marko95}

\begin{figure}[t!]
\includegraphics[scale=.30]{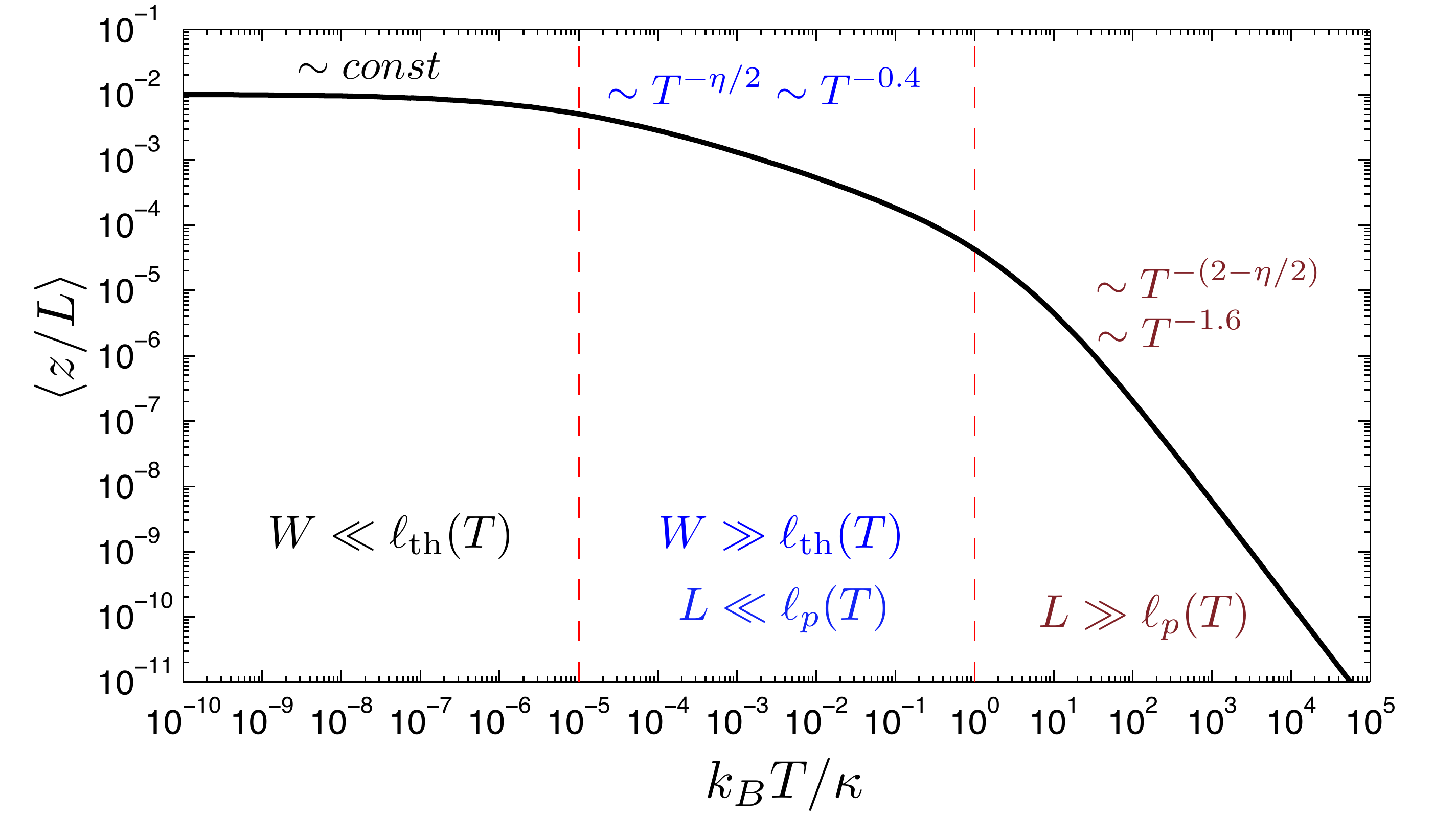}
\caption{(Color online) Response of ribbons (neglecting quantum fluctuations) to a small bending force at various temperatures for fixed $W$, $L$, $F$, $\kappa$ and $Y$. Three regimes appear for the parameter choices, $F L^2/3W\kappa=0.01$, $Y W^2/\kappa =10^5$, $L/W=10^{2}$.
}
\label{fig:bending_temp}
\end{figure}
To highlight the difference between conventional polymers and thermalized ribbons with $W \gg \ell_\textrm{th}$, consider the response of ribbons to a small bending force, $F L^2/W \kappa \ll 1$. Fig.~\ref{fig:bending_temp} shows results for a wide variety of temperatures, obtained by inserting temperature dependences hidden in $A_1$ and $\ell_p$. We find three distinct regimes:
 At small temperatures, where $W \ll \ell_\textrm{th}\sim \kappa/\sqrt{k_B T Y}$, thermal fluctuations are negligible and ribbon behaves like a classic cantilever with bare elastic parameters, $\langle z \rangle = F L^3/(3 \kappa W)$. As the temperature increases, the thermal length scale drops and eventually becomes smaller than the ribbon width ($\ell_\textrm{th} \ll W$). In this regime the renormalized bending rigidity is increased due to thermal fluctuations and the cantilever deflection is smaller $\langle z \rangle = F L^3 \ell_\textrm{th}^{\eta}/(3 \kappa W^{1 + \eta}) \sim T^{-\eta/2}$. As temperature increases even further, eventually the persistence length $\ell_p$ becomes smaller than the ribbon length $L$. As noted above in this semi-flexible regime the deflection now becomes $\langle z \rangle = 4 \kappa F L W^{1+\eta} /(3 (k_B T)^2 \ell_\textrm{th}^\eta) \sim T^{-(2-\eta/2)}$ and drops even faster with temperature, as the ribbon transforms from a cantilever into a random coil. Note that with rising temperatures the cutoff length scale $\ell_\sigma$ associated with ribbon tension [see Eq.~(\ref{eq:tension_length})] also increases, but never becomes relevant.

\begin{figure}[t!]
\includegraphics[scale=.30]{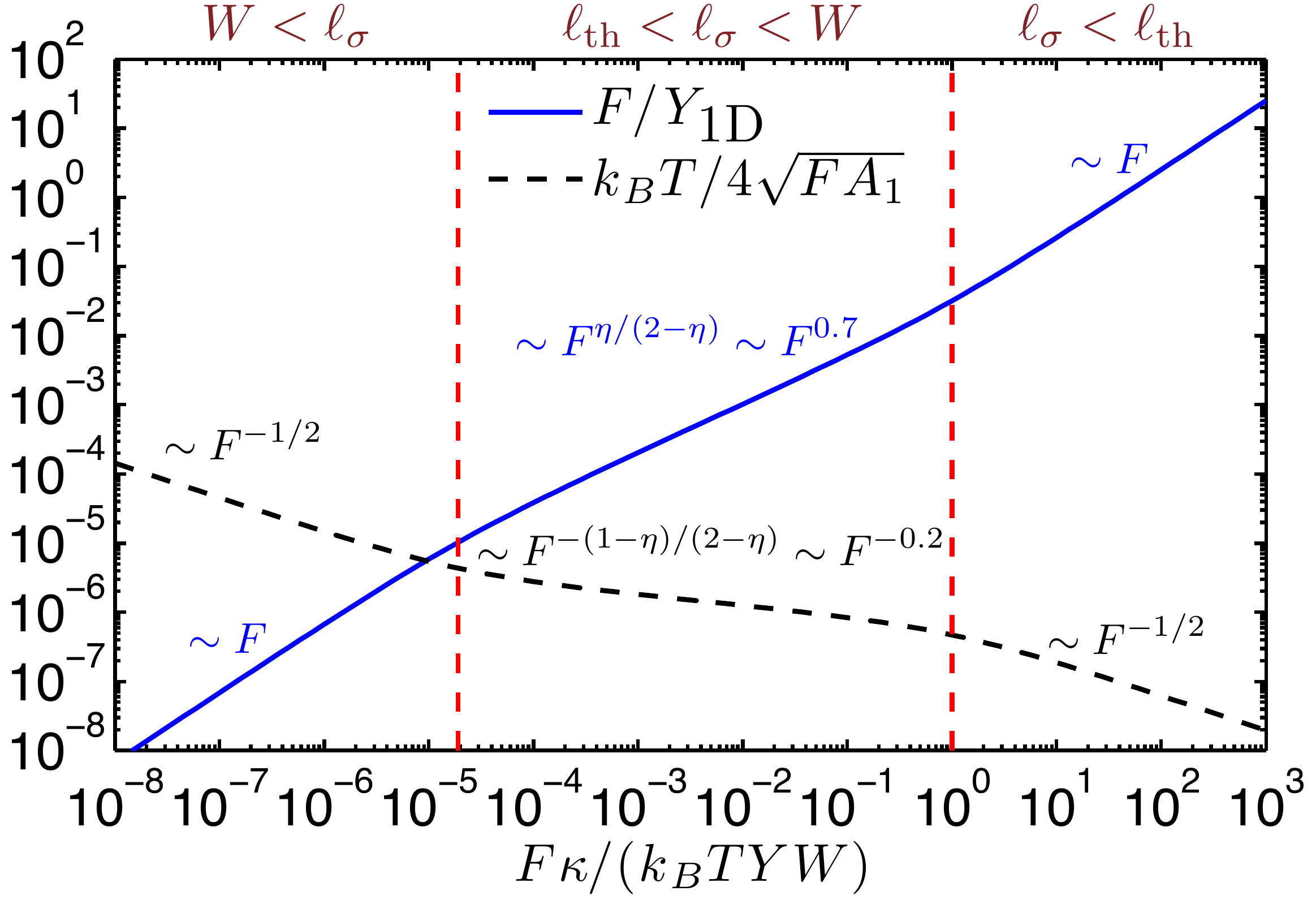}
\caption{(Color online) Contributions of backbone stretching (blue line) and entropic elasticity (dashed black line) describing the response to large ribbon pulling forces.
We chose parameters $k_B T/\kappa=1/40$ (suitable for graphene at room temperature), and $W/\ell_\textrm{th}=10^4$ (1 $\mu$m width ribbon at room temperature).
}
\label{fig:pulling_force}
\end{figure}

However, ribbons with \emph{large} pulling forces nevertheless show a non-trivial response due to the cutoff $\ell_\sigma$. For large pulling forces, $F \ell_p \gg k_B T$, we also need to include the stretching of the ribbon backbone, with the result similar to Ref.~\cite{moroz98}
$\left< \frac{z}{L} \right> \approx 1 + \frac{F}{Y_\textrm{1D}} - \frac{k_B T}{4 \sqrt{F A_1}}$ (see also supplemental materials),
where $Y_\textrm{1D}=W Y_R(W)$ is the effective one dimensional ribbon Young's modulus. The middle term describes stretching of the ribbon backbone, and the final correction corresponds to the entropic contribution from ribbon fluctuations. As $F=\sigma_{xx} W$ increases the cutoff length scale $\ell_\sigma$ [Eq.~(\ref{eq:tension_length})] drops and we find two crossovers, first when this length scale crosses the ribbon width $W$ and finally when it drops below the thermal length scale $\ell_\textrm{th}$ (see Fig.~\ref{fig:pulling_force}). Especially interesting is the intermediate force regime with $\ell_\textrm{th} \ll \ell_\sigma \ll W$, where we find that the ribbon backbone stretches as $F/Y_\textrm{1D} \sim F^{\eta/(2-\eta)}$, which generalizes to ribbons the result of Ref.~\cite{guitter89} for the nonlinear stretching of two dimensional membranes under a \emph{uniform} tension. See supplemental materials for further discussion.

For graphene ribbons, where $\ell_\textrm{th} \sim 1 \textrm{\AA}$ at room temperature, the experiments of Blees \emph{et al.}~\cite{blees15} on ribbons of width $W=10 \mu\textrm{m}$, confirmed a renormalized bending rigidity $\kappa_R(W)/\kappa \sim 6000$, consistent with Eq.~(\ref{eq:renormalized_elastic_constants}). The corresponding persistence length is of order of meters. Thus $\ell_p \gg L$ for graphene ribbons of lengths $L \sim 10-100 \mu\textrm{m}$, which should behave like conventional cantilevers with, however, a strongly renormalized $L$ independent bending rigidity. Probing the semi-flexible regime requires narrower ribbons of order 10 nanometers width, so that the persistence length should be in the experimentally accessible regime of 10-100 micrometers. Although the value of the critical pulling tension, beyond which thermal fluctuations become irrelevant, is $F/W =\sigma^* \sim 10\textrm{N}/\textrm{m}$ for graphene, one could observe interesting behavior for smaller tensions where $\ell_\textrm{th} < \ell_\sigma < W$. Additional novel behavior can arise for free-standing sheets at sufficiently high temperatures even when $L \approx W$. To see this, consider the correlation function of the membrane normals $\hat{\bf n}(x,y) = (-\partial_x f,-\partial_y f,1)/\sqrt{1 + |\nabla f|^2}$ that defines the flat phase~\cite{nelson87}. There is a power law approach to long range order,
$\left< \hat{\bf n}({\bf r}_a) \cdot \hat {\bf n}({\bf r}_b) \right> = 1 - \frac{k_B T}{2 \pi \kappa} \left[\eta^{-1} + \ln (\ell_\textrm{th}/a_0)\right] + C \frac{k_B T}{\kappa} \left(\frac{\ell_\textrm{th}}{|{\bf r}_a - {\bf r}_b|}\right)^\eta$,
where $C$ is a positive constant of order unity and $a_0$ is microscopic cutoff, of order the graphene lattice spacing (see supplemental materials which includes the effect of an isotropic external stress). The second term represents the reduction in the long range order due to thermal fluctuations. When this term becomes the same size as the first (i.e. for $k_B T \gtrsim 2 \pi \kappa \eta$), the low temperature flat phase should transform into a entropically dominated crumpled ball, with a size limited by self-avoidance, provided monolayer sheets such as graphene maintain their integrity~\cite{kantor87}. The transition temperature to isotropic crumpling could be lowered by creating a graphene sheet with a periodic array of holes or cuts. (Although cuts could be deployed with equal numbers at 120 degree angles,  an array of parallel cuts could lead to a system that is crumpled in one direction, but tube-like  in another, a situation studied theoretically in Ref.~\cite{radzihovsky95}.) While we have some understanding of force-free conformations~\cite{paczuski88}, little is known about the mechanical properties of free-standing membranes at or above this crumpling transition. There is evidence from computer simulations of a high temperature compact phase, where attractive van der Waals interactions are balanced by self-avoidance~\cite{abraham90}. We hope this paper will stimulate further investigations on these problems in the spirit of single-molecule experiments on linear polymers~\cite{bustamante03, moerner03}.

We acknowledge support by the National Science Foundation, through grants DMR1306367 and DMR1435999, and through the Harvard Materials Research and Engineering Center through Grant DMR-1420570. We would like to acknowledge conversations with P. McEuen, M. Blees, M. Bowick and R. Sknepnek.

\bibliography{library}

\newpage

\setcounter{equation}{0}
\renewcommand\theequation{S\arabic{equation}}

\setcounter{figure}{0}
\renewcommand\thefigure{S\arabic{figure}}

\begin{widetext}

\begin{center}
\huge{Supplementary Information}
\end{center}
\ \\
\normalsize

In this supplemental material we provide detailed calculations that were omitted in the main text for clarity. Section I describes the renormalization group treatment of membranes under tension. In Sec. II we present how pulling and bending of ribbons can be mapped to a time evolution of rotating top in quantum mechanics, and how tools from quantum mechanics can be used to calculate the force-extension curves for ribbons.

\section{Renormalization group treatment of membranes under tension}
Our goal is to analyze properties of fluctuating membranes under external tension $\sigma_{ij}$ with the renormalization group approach. The free energy cost of membrane deformations under tension is
\begin{equation}
E = \int\! d^2{\bf x} \, \frac{1}{2} \left[\lambda u_{ii}^2 + 2 \mu u_{ij}^2 + \kappa K_{ii}^2 - 2 \kappa_G \det(K_{ij})\right] - \oint \! ds \, \hat m_i \sigma_{ij} u_j,
\label{eq:plates}
\end{equation}
where $u_{ij}=(\partial_i u_j + \partial_j u_i + \partial_i f \partial_j f)/2$ is the nonlinear strain tensor, $K_{ij}=\partial_i \partial_j f$ is the bending strain tensor, the $u_i$ are in-plane deformations, $f$ is the out-of-plane deformation, and $\hat m_i$ describes a normal vector to the membrane boundary. Using the divergence theorem we can convert the boundary work term to the area integral, such that the free energy becomes
\begin{equation}
E = \int\! d^2{\bf x} \, \left( \frac{1}{2} \left[\lambda u_{ii}^2 + 2 \mu u_{ij}^2 + \kappa K_{ii}^2 - 2 \kappa_G \det(K_{ij}) \right]  - \sigma_{ij} u^0_{ij}\right),
\label{eq:free_energy}
\end{equation}
where $u_{ij}^0=(\partial_i u_j + \partial_j u_i)/2$ is the linear part of the strain tensor. Since the in-plane deformations $u_i$ only appears quadratically in (Eq.~\ref{eq:free_energy}), we can integrate them out to derive the effective free energy for the out-of-plane deformations,~\cite{nelsonB} 
\begin{eqnarray}
\frac{E}{A} &=& \sum_{\bf q} \frac{1}{2} \left[\kappa q^4 + \sigma_{ij} q_i q_j \right] f({\bf q}) f(-{\bf q}) \nonumber \\
&&+ \sum_{\substack{{\bf q}_1 + {\bf q}_2 = {\bf q} \ne {\bf 0}\\{\bf q}_3 + {\bf q}_4 = -{\bf q}\ne{\bf 0}}} \frac{Y}{8} 
\left[q_{1i} P_{ij}^T({\bf q}) q_{2j} \right] \left[q_{3i} P_{ij}^T({\bf q}) q_{4j} \right] f({\bf q}_1) f({\bf q}_2) f({\bf q}_3) f({\bf q}_4),
\label{eq:effective_free_energy}
\end{eqnarray}
where the Young's modulus is $Y = 4 \mu (\mu + \lambda)/(2 \mu + \lambda)$, the projection operator $P_{ij}^T({\bf q}) = \delta_{ij} - q_i q_j / q^2$, $A$ is the membrane area and the Fourier modes are $f({\bf q}) = \int (d^2{\bf r}/A) e^{-i {\bf q} \cdot {\bf r}} f({\bf r})$.
From the expression above we can clearly see that positive components of the membrane tension $\sigma_{ij}$ constrain the out-of-plane fluctuations $f$.

To implement a momentum shell renormalization group, we first integrate out all Fourier modes in a thin momentum shell $\Lambda/b < q < \Lambda$, where $\Lambda$ is microscopic cutoff and $b\equiv \ell \Lambda=e^{\delta}$ with $\delta \ll 1$. Next we rescale lengths and fields~\cite{aronovitz88, radzihovsky91}
\begin{eqnarray}
{\bf x} &=& b {\bf x}', \nonumber\\
f({\bf x}) & = & b^{\zeta_f} f'({\bf x}').
\end{eqnarray}
We find it convenient to work directly with a $D=2$ dimensional membrane embedded in $d=3$ space, rather than introducing an expansion in $\epsilon = 4 -D$.~\cite{aronovitz88} 
Finally, we define new elastic constants $\kappa'$, $Y'$, and external tension $\sigma_{ij}'$, such that the free energy functional in Eq.~(\ref{eq:effective_free_energy}) retains the same form after the first two renormalization group steps. It is common to introduce $\beta$ functions~\cite{amitB}, which define the flow of elastic constants
 \begin{eqnarray}
 \beta_\kappa & = & \frac{\partial \kappa'}{\partial \ln b} = 2 (\zeta_f - 1) \kappa + Z_\kappa , \nonumber \\
 \beta_Y & = & \frac{\partial Y'}{\partial \ln b} = 2 (2 \zeta_f - 1) Y + Z_Y, \nonumber \\
 \beta_{ij} & = & \frac{\partial \sigma'_{ij}}{\partial \ln b} = 2 \zeta_f \sigma_{ij}.
 \end{eqnarray} 
 \begin{figure}[t!]
\includegraphics[scale=.7]{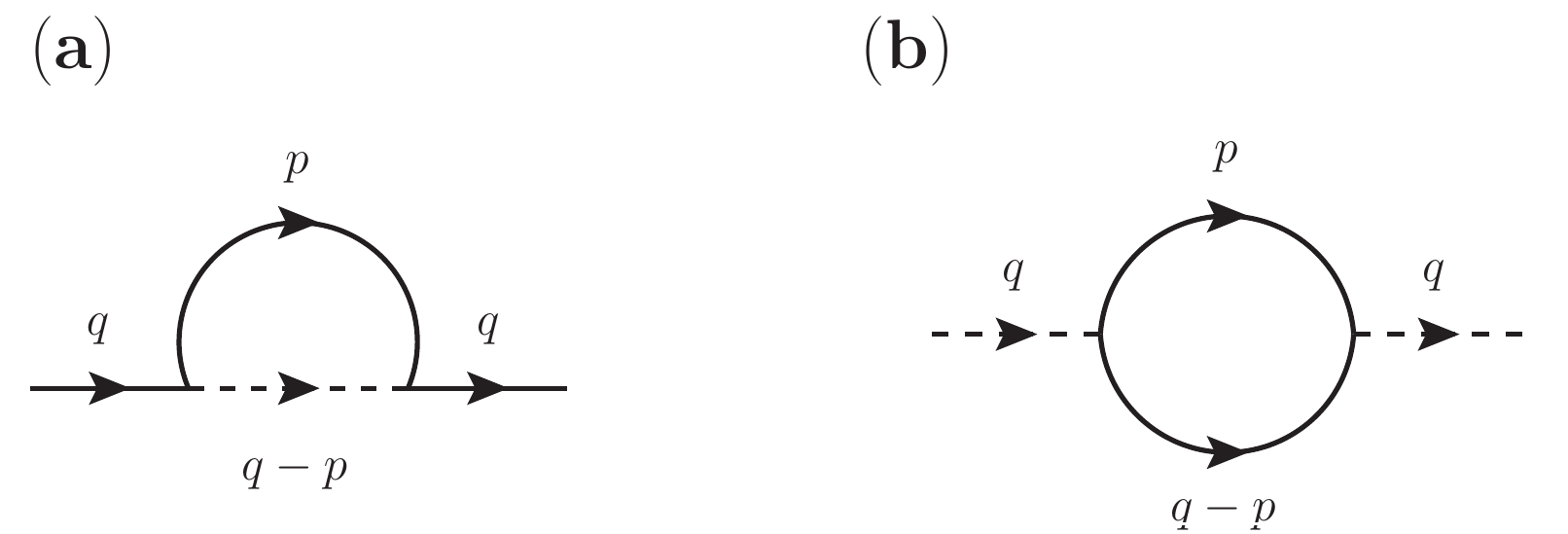} 
\caption{One loop corrections to the renormalization of (a) $\kappa$ and (b) $Y$. Solid lines represent propagators for the out-of-plane displacements $f({\bf q})$ and dashed lines represent the momentum carried by the vertex $Y$.
}
\label{fig:diagrams}
\end{figure}
Above we introduced $Z$ functions, which result from the integrals of modes over the momentum shell. To one loop order (see Fig.~\ref{fig:diagrams}), the $Z$ functions read
\begin{eqnarray}
Z_\kappa & = & + \frac{\partial}{\partial \ln b} \left(Y \sum_{\frac{\Lambda}{b} < p < \Lambda}  \left(1 - ({\hat {\bf q}} \cdot \hat {\bf p})^2 \right)^2 \left< f({\bf p}) f(-{\bf p}) \right>  \right), \nonumber \\
Z_Y & = & - \frac{\partial}{\partial \ln b} \left( \frac{Y^2 A}{2 k_B T}\sum_{\frac{\Lambda}{b} < p < \Lambda}  \left(1 - ({\hat {\bf q}} \cdot \hat {\bf p})^2 \right)^2 p^4 \left< f({\bf p}) f(-{\bf p}) \right>^2  \right),
\label{eq:Z}
\end{eqnarray}
where $\Lambda$ is the microscopic momentum cutoff and $A$ is the undeformed membrane area. Note that the only change in the stress tensor $\sigma_{ij}$ to this order arises from the rescaling factor $\zeta_f$.
Upon assuming that the initial membrane tension $\sigma_{ij}$ is small, such that $\sigma_{ij} \ll \kappa \Lambda^2$, then $\left<f(p)f(-p)\right> \approx k_B T / (A \kappa p^4)$ in equations above and the $\beta$ functions in one loop approximation become
\begin{eqnarray}
 \beta_\kappa & = & 2 (\zeta_f - 1) \kappa + \frac{3 Y k_B T}{16 \pi \kappa \Lambda^2} , \nonumber \\
 \beta_Y & = & 2 (2 \zeta_f - 1) Y - \frac{3 Y^2 k_B T}{32 \pi \kappa^2 \Lambda^2}, \nonumber \\
 \beta_{ij} & = &  2 \zeta_f \sigma_{ij}.
\end{eqnarray}
It is convenient to chose $\zeta_f$ such that $\beta_\kappa=0$, which results in $\zeta_f = 1 - \frac{3 Y k_B T}{32 \pi \kappa^2 \Lambda^2}$, and
\begin{eqnarray}
 \beta_Y & = & 2 Y - \frac{15 Y^2 k_B T}{32 \pi \kappa^2 \Lambda^2}, \nonumber \\
 \beta_{ij} & = & 2 \left(1 - \frac{3 Y k_B T}{32 \pi \kappa^2 \Lambda^2} \right) \sigma_{ij}.
\end{eqnarray}
By repeating the renormalization group procedure, we integrate out modes at the smallest length scale and evolve the Young's modulus $Y$ and external tension $\sigma_{ij}$. Initially, they both grow rapidly
\begin{eqnarray}
Y(\ell) &\approx& Y \times (\ell \Lambda)^2, \nonumber \\
\sigma_{ij} (\ell) & \approx & \sigma_{ij} \times (\ell \Lambda)^2,
\end{eqnarray}
where we integrated out all modes on scales smaller than $\ell$. Once we integrate out all modes up to the scale $\ell_\textrm{th} \sim \kappa/\sqrt{k_B T Y}$ Young's modulus reaches a fixed point
\begin{eqnarray}
Y^* = \frac{64 \pi \kappa^2 \Lambda^2}{15 k_B T} \sim Y \times (\ell_\textrm{th} \Lambda)^2.
\end{eqnarray}
At the fixed point we introduce the exponent $\eta$, such that $\zeta_f = 1 - \eta/2$. Note that $\zeta_f \approx 1$ initially, before we reach the fixed point. In the one loop approximation we find $\eta=4/5$, which approximates the value of $\eta \approx 0.82$ obtained by the self-consistent screening approximation~\cite{ledoussal92} and $\eta \approx 0.85$ obtained by the non-perturbative renormalization group calculations~\cite{kownacki09}. This result differs from a formal one loop $\epsilon=4-D$ expansion, which results in $\eta=12 \epsilon/25$,~\cite{aronovitz88} because we have performed the one loop calculations directly for $D=2$ dimensional membranes, rather than calculating them for small $\epsilon$, i.e. for $D\approx 4$ dimensional membranes.

By continuing with the renormalization group procedure and integrating out modes beyond the scale $\ell_\textrm{th}$, we find that the initially small membrane tension now grows as
\begin{equation}
\sigma_{ij} (\ell > \ell_\textrm{th}) = \sigma_{ij}\times (\ell/\ell_\textrm{th})^{2-\eta}\times (\ell_\textrm{th} \Lambda)^2.
\end{equation}
Eventually, the membrane tension becomes large enough that it becomes important. This happens at the scale
\begin{equation}
\ell_\sigma \sim \left( \frac{\kappa}{\sigma \ell_\textrm{th}^\eta} \right)^{1/(2-\eta)},
\end{equation}
when $\sigma_{ij} (\ell_\sigma) \sim \kappa \Lambda^2$. At this stage, we have to take into account the membrane tension, when evaluating the $Z$ functions in Eq.~(\ref{eq:Z}). In next subsections, we describe what happens for membranes under various external tension conditions. We first discuss membranes with $W \sim L$, and then move on to discuss ribbons with $L \gg W$.

\subsection{Membranes under uniform tension}
We first consider membranes under uniform tension $\sigma_{ij} = \sigma \delta_{ij}$. After integrating modes on scales smaller than $\ell_\sigma$, the membrane tension becomes relevant and beyond this point we can approximate $\left<f(p)f(-p)\right> \approx k_B T / (A \sigma p^2)$ in Eqs.~(\ref{eq:Z}). With this change, the $\beta$ functions become
\begin{eqnarray}
 \beta_\kappa & = & 2 (\zeta_f - 1) \kappa + \frac{3 Y k_B T}{16 \pi \sigma} , \nonumber \\
 \beta_Y & = & 2 (2 \zeta_f - 1) Y - \frac{3 Y^2 k_B T \Lambda^2}{32 \pi \sigma^2}, \nonumber \\
 \beta_{ij} & = &  2 \zeta_f \sigma_{ij}.
\end{eqnarray}
It is now convenient to set $\zeta_f=0$ so that the uniform tension remains unchanged. We then find that both the bending rigidity $\kappa$ and the Young's modulus $Y$ flow to 0 at large length scales,
\begin{eqnarray}
\kappa (\ell > \ell_\sigma) & \sim & \kappa \times (\ell / \ell_\sigma)^{-2}, \nonumber \\
Y (\ell > \ell_\sigma) & \sim & Y^* \times (\ell / \ell_\sigma)^{-2}.
\end{eqnarray}
In this regime the external tension dominates and thermal fluctuations are unimportant. After rescaling lengths and fields back to the initial units we find that the height correlation function for the out-of-plane flexural phonons is
\begin{eqnarray}
\left<f({\bf q}) f(-{\bf q}) \right> =  \left\{
\begin{array}{c l}
k_B T/(A \kappa q^4), & q > \ell_\textrm{th}^{-1} \\
k_B T/(A \kappa q^{4-\eta} \ell_\textrm{th}^{-\eta}), & \ell_\textrm{th}^{-1}  >  q> \ell_\sigma^{-1} \\
k_B T/(A \sigma q^2), &  \ell_\sigma^{-1} > q
\end{array}
\right.
\equiv \frac{k_B T}{A (\sigma q^2 + \kappa_R(q) q^4)}
\end{eqnarray}
It is useful to calculate how the membrane expands due to the external tension in the presence of thermal fluctuations. We find that the fractional area change is
\begin{eqnarray}
\frac{\delta A}{A} &=& \left< u_{ii}^0 \right> =  \frac{\sigma}{(\mu+\lambda)} - \frac{1}{2} \sum_{{\bf q}} q^2 \left<f({\bf q}) f(-{\bf q}) \right>, \nonumber \\
\frac{\delta A}{A} & \approx & - \frac{k_B T}{4 \pi \kappa} \left[\frac{1}{\eta} + \ln(\ell_\textrm{th} \Lambda) \right]  + \frac{k_B T}{4 \pi \kappa} \left[\eta^{-1}- \frac{1}{2} \right] \left( \frac{\kappa \sigma}{k_B T Y}\right)^{\eta/(2-\eta)} + \frac{\sigma}{(\mu+\lambda)}.
\label{eq:uniform_stretching}
\end{eqnarray}
The first term above corresponds to the shrinking of membrane due to thermal fluctuations, and reflects a negative coefficient of thermal expansion when $\sigma=0$,
\begin{equation}
\alpha = \frac{1}{A} \frac{dA}{dT} \approx - \frac{k_B}{4 \pi \kappa} \left[\frac{1}{\eta} - \frac{1}{2} + \ln(\ell_\textrm{th} \Lambda) \right].
\end{equation}
The second term describes the nonlinear stretching for small uniform tension $\sigma$ in the presence of thermal fluctuations~\cite{guitter89}. The last term is the conventional linear response result, describing membrane stretching in the absence of thermal fluctuations. This last term only dominates for large tensions, $\sigma \gtrsim k_B T Y/\kappa$, which corresponds to $\ell_\textrm{th} \gtrsim \ell_\sigma$.

Finally, we present the correlation function of the membrane normals $\hat{\bf n}(x,y) = (-\partial_x f,-\partial_y f,1)/\sqrt{1 + |\nabla f|^2}$ that defines the flat phase~\cite{nelson87}. When deformations are small the correlation function of the membrane normals is approximately
\begin{equation}
\left< \hat {\bf n}({\bf r}_a) \cdot \hat {\bf n}({\bf r}_b) \right> \approx 1 - \sum_{\bf q} q^2 \left[1 - e^{i {\bf q} \cdot ({\bf r}_a - {\bf r}_b)} \right] \left<|f({\bf q})|^2 \right>.
\end{equation}
For small tension  $\sigma \lesssim k_B T Y/\kappa$ this correlation function evaluates to
\begin{eqnarray}
\left< \hat {\bf n}({\bf r}_a) \cdot \hat {\bf n}({\bf r}_b) \right> &\approx& 1 - \frac{k_B T}{(2 \pi \kappa)} \left[\eta^{-1} + \ln (\ell_\textrm{th} \Lambda)\right] + \frac{k_B T}{(2 \pi \kappa)} (\eta^{-1}-2^{-1}) \left(\frac{\kappa \sigma}{k_B T Y} \right)^{\eta/(2 -\eta)}  \nonumber \\
&&
+\frac{k_B T}{\kappa}  \left\{
\begin{array}{l l}
C  \left( \frac{\ell_\textrm{th}}{|{\bf r}_a - {\bf r}_b|}\right)^\eta, & \ell_\textrm{th} \ll |{\bf r}_a - {\bf r}_b| \ll \ell_\sigma \\
D \left( \frac{\ell_\textrm{th}}{\ell_\sigma}\right)^\eta  e^{-|{\bf r}_a - {\bf r}_b|/\ell_\sigma} ,& \ell_\sigma \ll |{\bf r}_a - {\bf r}_b|
\end{array}
\right.,
\end{eqnarray}
where $C = \frac{1}{2 \pi} \int_0^\infty \frac{dx}{x^{1-\eta}} J_0(x) \approx 0.2$, $D$ is another constant of order unity and $J_0(x)$ is the Bessel function of the first kind. The second term in the equation above represents the reduction in the long range order between normals due to thermal fluctuations and the third term shows how this long range order is restored with external tension. For large tension  $\sigma \gtrsim k_B T Y/\kappa$, where the effects of thermal fluctuations are suppressed, we find
\begin{equation}
\left< \hat {\bf n}({\bf r}_a) \cdot \hat {\bf n}({\bf r}_b) \right> \approx 1 - \frac{k_B T}{(4 \pi \kappa)} \ln \left[1 + \frac{\kappa \Lambda^2}{\sigma}\right] + \frac{k_B T}{(2 \pi \kappa)} K_0 \left({|{\bf r}_a - {\bf r}_b|} \sqrt{\sigma/\kappa}  \right),
\end{equation}
where $K_0(x)$ is the modified Bessel function of the second kind, which asymptotically scales as $K_0 (x) \asymp \sqrt{\pi/(2x)} e^{-x}$.

Note that the nonlinear dependence of the membrane extension $\left<u_{ii}^0\right>$ on the external tension $\sigma$ can be obtained simply from the scaling arguments. Since the external tension $\sigma$ is a conjugate variable to $\partial_j u_i$, their rescalings are connected. Once we rescale lengths as $x = b x'$ and in-plane deformations $u_i = b^{\zeta_u} u_i'$, then the external tension rescales as $\sigma = b^{\zeta_\sigma} \sigma'$ with $\zeta_\sigma=1 - D - \zeta_u$, where $D=2$ is the membrane dimensionality. We also know that the Ward identities associated with rotational symmetry connect rescaling of the in-plane and out-of-plane deformations such that $\zeta_u=2\zeta_f - 1$ and therefore $\zeta_\sigma = -2 \zeta_f$~\cite{guitter89}. As mentioned above we can extract exponent $\eta$ from $\zeta_f=1-\eta/2$, which leads to $\zeta_u=1-\eta$ and $\zeta_\sigma=-2+\eta$. Now we have all necessary ingredients to calculate the scaling of membrane extension as
\begin{eqnarray}
\left< \delta u_{ii}^0 (\sigma) \right> = \left< {\delta u_{ii}^0}' (\sigma') \right> b^{\zeta_u - 1} = \left< {\delta u_{ii}^0} (\sigma b^{-\zeta_\sigma}) \right> b^{\zeta_u - 1}.
\end{eqnarray}
Since the rescaling factor $b$ is arbitrary, we can pick $b = \sigma^{1/\zeta_\sigma}$ to find
\begin{equation}
\left< \delta u_{ii}^0 (\sigma) \right> = \left< \delta u_{ii}^0 (1) \right> \sigma^{(\zeta_u - 1)/\zeta_\sigma} =\textrm{const.} \times \sigma^{\eta/(2-\eta)}.
\end{equation}
Thus we found the same nonlinear scaling between the membrane stretching and the uniform tension as in Eq.~(\ref{eq:uniform_stretching}), which holds for small uniform tension.

\subsection{Membranes under uniaxial tension}
In this section we consider membranes under uniaxial tension $\sigma_{xx}>0$, while $\sigma_{yy}=\sigma_{xy}=0$. Upon again integrating out modes on scales smaller than $\ell_\sigma$, the membrane tension becomes important and beyond this point we have to take $\left<f(p)f(-p)\right> \approx k_B T / [A (\sigma_{xx} p_x^2 + \kappa p_y^4)]$ in Eqs.~(\ref{eq:Z}). Although we can ignore a term $\kappa (p_x^4+ 2 p_x^2 p_y^2)$ compared to $\sigma_{xx} p_x^2$, we have to keep the term with $\kappa p_y^4$. Once we integrate out modes from a thin shell $\Lambda/b < p < \Lambda$, we find that the quadratic term in the free energy becomes
\begin{eqnarray}
\frac{1}{2}\sum_{\bf q} f({\bf q}) f(-{\bf q}) \left\{ \sigma_{xx} q_x^2 + \kappa q_y^4 + \frac{Y k_B T \ln b}{4 \pi} \left[\frac{2 q_x^4}{\Lambda \sqrt{\kappa \sigma}} + \frac{1}{\sigma} \left(-3 q_x^4 + 6 q_x^2 q_y^2 + q_y^4 \right) \right] \right\}.
\end{eqnarray}
All new generated terms that involve $q_x$ are negligible compared to the $\sigma_{xx} q_x^2$. Therefore we can keep only the last term with $q_y^4$ to calculate the $\beta_\kappa$ function that renormalizes the bending rigidity,
\begin{equation}
 \beta_\kappa =  2 (\zeta_f - 1) \kappa + \frac{Y k_B T}{4 \pi \sigma}.
\end{equation}
For the quartic term with momentum-dependent Young's modulus $Y$ we find that after the momentum shell integration there are again anisotropic contributions in terms of $q_x$ and $q_y$. Significantly, all renormalizations of $Y$  are negative and $\beta$ functions now take the form
\begin{eqnarray}
 \beta_\kappa &= & 2 (\zeta_f - 1) \kappa + \frac{Y k_B T}{4 \pi \sigma}, \nonumber \\
 \beta_Y & = & 2 (2 \zeta_f - 1) Y - |Z_Y|, \nonumber \\
 \beta_{ij} & = &  2 \zeta_f \sigma_{ij}.
\end{eqnarray}
As for the uniform tension case, we choose $\zeta_f=0$ to fix the uniaxial tension, and find that both the bending rigidity $\kappa$ and the Young's modulus $Y$ again flow to 0 as
\begin{eqnarray}
\kappa (\ell > \ell_\sigma) & \sim & \kappa \times (\ell / \ell_\sigma)^{-2}, \nonumber \\
Y (\ell > \ell_\sigma) & \sim & Y^*  \times (\ell / \ell_\sigma)^{-2}.
\end{eqnarray}
After rescaling lengths and fields back to the initial units we find that the correlation function for the out-of-plane deformations becomes highly anisotropic
\begin{eqnarray}
\left<f({\bf q}) f(-{\bf q}) \right> =  \left\{
\begin{array}{c l}
k_B T/(A \kappa q^4), & q > \ell_\textrm{th}^{-1} \\
k_B T/(A \kappa q^{4-\eta} \ell_\textrm{th}^{-\eta}), & \ell_\textrm{th}^{-1}  >  q> \ell_\sigma^{-1} \\
k_B T/(A [\sigma_{xx} q_x^2 + \kappa q_y^4 (\ell_\sigma/\ell_\textrm{th})^{\eta} ]), &  \ell_\sigma^{-1} > q
\end{array}
\right.
\equiv \frac{k_B T}{A (\sigma_{xx} q_x^2 + \kappa_R(q) q^4)}.
\label{eq:corr}
\end{eqnarray}
We can now use this result to calculate the membrane strains associated with uniaxial stretching
\begin{eqnarray}
 \left< u_{xx}^0 \right> &=&  \frac{\sigma_{xx}}{Y} - \frac{1}{2} \sum_{{\bf q}} q_x^2 \left<f({\bf q}) f(-{\bf q}) \right>, \nonumber \\
 \left< u_{xx}^0 \right> &\approx& - \frac{k_B T}{8 \pi \kappa} \left[\eta^{-1} + \ln(\ell_\textrm{th} \Lambda) \right]  + \frac{k_B T}{8 \pi \kappa} \left[\eta^{-1}- 1 + \sqrt{2}-\sinh^{-1}(1) \right] \left( \frac{\kappa \sigma_{xx}}{k_B T Y}\right)^{\eta/(2-\eta)} + \frac{\sigma_{xx}}{Y}, \nonumber \\
 \left< u_{yy}^0 \right> &=&  - \frac{\nu \sigma_{xx}}{Y} - \frac{1}{2} \sum_{{\bf q}} q_x^2 \left<f({\bf q}) f(-{\bf q}) \right>, \nonumber \\
 \left< u_{yy}^0 \right> &\approx& - \frac{k_B T}{8 \pi \kappa} \left[\eta^{-1} + \ln(\ell_\textrm{th} \Lambda) \right]  + \frac{k_B T}{8 \pi \kappa} \left[\eta^{-1} +1 - \sqrt{2}-\sinh^{-1}(1) \right] \left( \frac{\kappa \sigma_{xx}}{k_B T Y}\right)^{\eta/(2-\eta)} - \frac{\nu \sigma_{xx}}{Y}, \nonumber \\
\label{eq:uxx}
\end{eqnarray}
where $\nu = \lambda/(2 \mu + \lambda)$ is the two-dimensional Poisson ratio.
Again, the first terms in the second and fourth lines describe membrane shrinkage due to thermal fluctuations, the second terms correspond to nonlinear membrane stretching in the presence of thermal fluctuations, and the last terms correspond to the zero temperature response, which becomes relevant for  $\sigma_{xx} \gtrsim k_B T Y/\kappa$.  The power law scalings above are accurate, but the numerical prefactors are approximate. In order to calculate numerical prefactors exactly, we would need to know how the correlation function in Eq.~(\ref{eq:corr}) behaves in transition regions. In principle, the renormalized Poisson's ratio is calculated as
\begin{eqnarray}
\nu_R = -\frac{ \left<\delta u_{yy}^0 \right>}{ \left<\delta u_{xx}^0 \right>},
\end{eqnarray}
where $\left<\delta u_{ij}^0 \right>$ describes the relative change, when the uniaxial tension is increased from zero to $\sigma_{xx}$.
Because our numerical prefactors in Eqs.~(\ref{eq:uxx}) are just approximate we cannot determine the precise value of the renormalized Poisson's ratio $\nu_R$ in the regime dominated by thermal fluctuations, but we know that the $\nu_R$ transitions to the zero temperature value $\nu$ for large pulling tension, i.e. $\sigma_{xx} \gtrsim k_B T Y/\kappa$.

\subsection{Pulling of ribbons}
Finally, we comment on pulling on large aspect ratio ribbons of length $L$ and width $W \ll L$. After integrating out all degrees of freedom on scales smaller than the width $W$, the resulting strain tensors $u_{ij}$ and $K_{ij}$ depend only on the $x$ coordinate and the renormalized elastic constants are evaluated at $q=2\pi/W$. This results in an effectively one dimensional free energy model for the ribbon
\begin{equation}
E = \int_0^L\! d{x} \, W \left( \frac{1}{2} \left[\lambda_R u_{ii}^2 + 2 \mu_R u_{ij}^2 + \kappa_R K_{ii}^2 - 2 \kappa_{GR} \det(K_{ij}) \right]  -  \sigma_{xx} u^0_{xx}\right).
\label{eq:si:1d_free_energy}
\end{equation}
It is convenient to rewrite the effective free energy above in terms of one-dimensional Fourier variables $s(q)\equiv\int (dx/L) e^{-i q x} s(x)$ and to separate out the uniform strain $u_{ij}^0$.~\cite{nelsonB} The resulting free energy reads
\begin{eqnarray}
\frac{E}{WL} &=& \frac{1}{2} \left[\lambda_R \left( u^0_{ii} \right)^2 + 2 \mu_R \left( u^0_{ij} \right)^2 \right]+ \frac{1}{2} \sum_q \left[\kappa_R q^4 |f(q)|^2
 + (2 \mu_R + \lambda_R) q^2 |u_x(q)|^2 + \mu_R q^2 |u_y (q)|^2 \right] - \sigma_{xx} u^0_{xx} \nonumber\\
 && + \frac{1}{2} \sum_q \left[ \lambda_R u_{ii}^0 + 2 \mu_R  u^0_{xx} \right] q^2 |f(q)|^2
 + \frac{i}{2} \sum_{q_1, q_2} (2 \mu_R + \lambda_R) q_1 q_2 (q_1 + q_2) u_x(q_1) f(q_2) f(-q_1-q_2) \nonumber \\
 && - \frac{(2 \mu_R + \lambda_R)}{8} \sum_{q_1, q_2, q_3} q_1 q_2 q_3 (q_1 + q_2 + q_3) f(q_1) f(q_2) f(q_3) f(-q_1 - q_2 - q_3).
 \end{eqnarray}
Because the in-plane deformations $u_y(q)$ decouple the shear modulus $\mu_R$ does not get further renormalized. Similarly, we find that the bending rigidity $\kappa_R$ does not get further renormalized. To see this, we integrate out the in-plane modes $\{u^0_{ij},u_i(q)\}$ to derive the effective free energy
\begin{equation}
\frac{F}{LW} = \frac{1}{2} \sum_q \left[ \kappa_R q^4 + \sigma_{xx} q^2 \right] |f(q)|^2.
\end{equation}
However, the in-plane modulus $2\mu_R + \lambda_R$ associated with the in-plane deformation $u_x(q)$ suffers significant renormalizations. This can be shown with the momentum shell renormalization group by integrating out all Fourier modes in a thin momentum shell $\Lambda/b < q < \Lambda$ and rescaling lengths and fields as
\begin{eqnarray}
x&=& b x', \nonumber \\
u_i(x) & = & b^{\zeta_u}  u_i' (x'), \nonumber \\
f(x) & = & b^{\zeta_f} f'(x'), \nonumber \\
\sigma_{xx} & = & b^{\zeta_\sigma} \sigma_{xx}'.
\end{eqnarray}
Note that the momentum cutoff is now $\Lambda=2 \pi/W$, because we already integrated out all degrees of freedom on scales smaller than $W$. As in previous sections we define $\beta$ functions that dictate the flow of elastic constants
\begin{eqnarray}
\beta_\kappa &=& \frac{\partial \kappa'}{\partial \ln b} = 2 (\zeta_f - 1) \kappa, \nonumber \\
\beta_\mu &=&  \frac{\partial \mu'}{\partial \ln b} =2 \zeta_u \mu, \nonumber \\
\beta_{2 \mu + \lambda} &=& \frac{\partial (2 \mu + \lambda)'}{\partial \ln b} = 2 \zeta_u (2 \mu + \lambda) - Z_{2 \mu + \lambda}, \nonumber \\
\beta_\sigma & = & \frac{\partial \sigma_{xx}'}{\partial \ln b} = -\zeta_\sigma \sigma_{xx}.
\end{eqnarray}
The Ward identities associated with rotational symmetry connect rescaling of the in-plane and out-of-plane deformations such that $\zeta_u=2\zeta_f - 1$~\cite{guitter89} and $\zeta_\sigma = -1 - \zeta_u$, because $\sigma_{xx}$ and $\partial_x u_x$ are conjugate variables.
 \begin{figure}[t!]
\includegraphics[scale=.7]{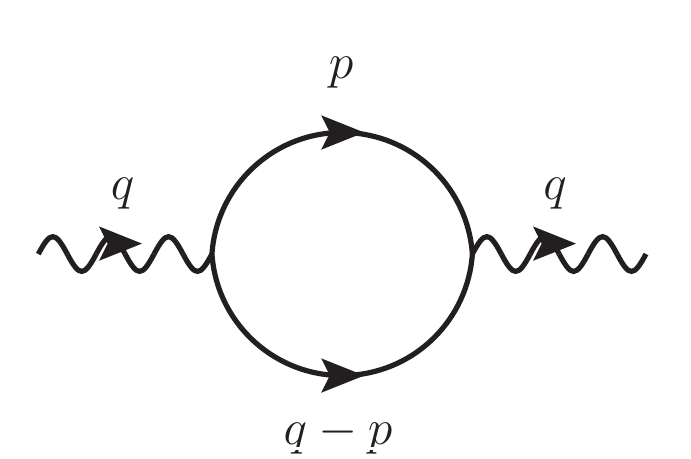} 
\caption{One loop corrections to the renormalization of  $2 \mu + \lambda$. Here, the solid and wiggly lines represent propagators for the out-of-plane displacement $f(q)$ and for the in-plane displacement $u_x(q)$, respectively.
}
\label{fig:diagrams2}
\end{figure}
The integrals of modes over the momentum shell now only affect the in-plane modulus $2\mu + \lambda$ and to one loop order (see Fig.~\ref{fig:diagrams2}) we find
\begin{equation}
Z_{2 \mu + \lambda} = \frac{\partial}{\partial \ln b} \left[ \frac{A (2 \mu + \lambda)^2}{2 k_B T} \sum_{\frac{\Lambda}{b} < p < \Lambda} p^4 \langle f(p) f(-p) \rangle^2 \right].
\end{equation}
Upon assuming that the external tension is small, i.e. $\sigma_{xx}  \ll \kappa_R \Lambda^2$, and choosing $\zeta_f=1$ to fix the bending rigidity $\kappa$, the flow of elastic constants is described by the $\beta$-functions,
\begin{subequations}
\begin{eqnarray}
\beta_{2 \mu + \lambda} & = & 2 (2 \mu + \lambda) - \frac{(2 \mu + \lambda)^2 k_B T}{2 \pi \kappa^2 \Lambda^3 W},  \label{eq:si:beta} \\
\beta_\sigma & = & 2 \sigma_{xx}.
\end{eqnarray}
\end{subequations}
Note that in the equations above the width $W$ also gets rescaled according to $W \rightarrow W/(\ell \Lambda)$. If the in-plane modulus $2 \mu + \lambda$ was small, then we would expect it to grow as 
\begin{equation}
2 \mu(\ell) + \lambda(\ell)  \sim (2 \mu_R + \lambda_R) \times (\ell \Lambda)^2.
\end{equation}
This modulus would keep growing until we integrate out all degrees of freedom up to the scale 
\begin{equation}
\ell^* \sim \left( \frac{\kappa_R^2 W}{k_B T (2 \mu_R + \lambda_R)}\right)^{1/3},
\end{equation}
where the second term in the $\beta_{2\mu+\lambda}$ function in Eq.~(\ref{eq:si:beta}) becomes relevant. However, for small tension $\sigma_{xx} \ll \kappa_R/W^2$, we find $\ell^* \sim W$, because the elastic moduli above have already suffered large renormalizations out to the scale $W$. Therefore the second term in Eq.~(\ref{eq:si:beta}) has to be taken into account immediately and the in-plane modulus flows as
\begin{equation}
2 \mu(\ell) + \lambda(\ell)  \sim (2 \mu_R + \lambda_R) \times (W/\ell)
\end{equation}
for $\ell \gg W$. This modulus keeps dropping until the external tension becomes relevant at scale $\ell_\sigma \sim \sqrt{\kappa_R/\sigma_{xx}}$. As in previous subsections the external tension introduces a cut-off length scale for the renormalization of the elastic modulus. By rescaling lengths and fields back to the original units we find the in-plane correlation function of displacements along the ribbon axis,
\begin{equation}
\langle u_x(q) u_x(-q) \rangle = \frac{k_B T}{L W (2 \mu_R + \lambda_R)} \left\{
\begin{array}{c c}
1/(q^5 W^3), & W \ll q^{-1} \ll \ell_\sigma \\
\ell_\sigma^3 /(q^2 W^3),& \ell_\sigma \ll q^{-1} \\
\end{array}
\right. .
\end{equation}
Note that for small external tension the renormalization produces large in-plane fluctuations $u_x$, suggesting that the description for the effective one dimensional free energy in Eq.~(\ref{eq:si:1d_free_energy}), which assumes small deformations about an approximately flat ribbon geometry, must eventually break down. In the next section we discuss how to treat ribbons with large deformations.

\section{Force-extension curve of ribbons due to thermal fluctuations}

Consider a long thin ribbon of length $L$, thickness $h$ (atomically thin for graphene!) and width $W$ in which we embed a position-dependent orthonormal triad frame $\{{\bf e}_1(s),{\bf e}_2(s),{\bf e}_3(s) \}$. Here $s\in [0,L]$ is an arclength coordinate along the ribbon midline, ${\bf e}_3$ is a unit tangent vector along this backbone, and ${\bf e}_1$ and ${\bf e}_2$ are unit normal vectors to the backbone as sketched below.
\begin{center}
\includegraphics[scale=.6]{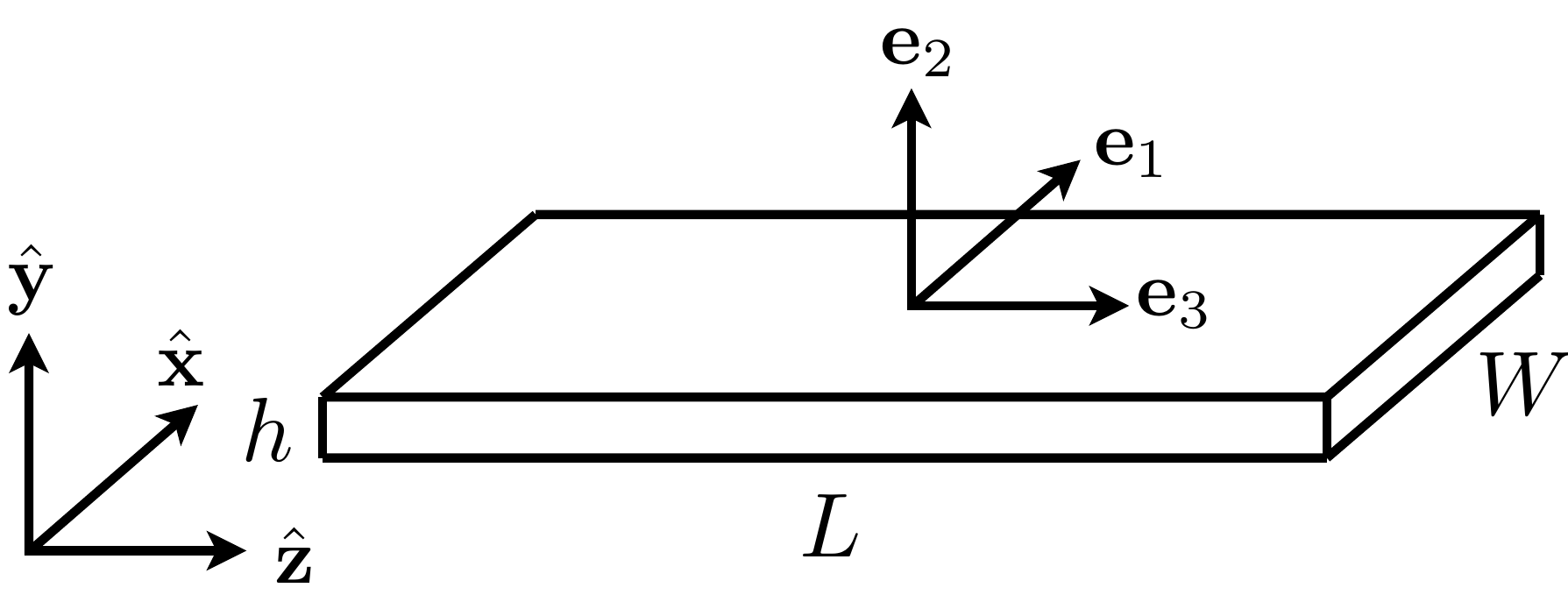} 
\end{center}
One way to express the rotation matrix $R$, which rotates the fixed laboratory frame $\{\hat {\bf x},\hat {\bf y}, \hat {\bf z}\}$ to the ribbon frame is to use Euler angles ${\bf \Theta}=\{\phi,\theta,\psi\}$,~\cite{landauBquantummechanics} via the decomposition $R({\bf \Theta})\equiv R_z(-\psi) R_y(-\theta) R_z(-\phi)$, such that $\{{\bf e}_1,{\bf e}_2,{\bf e}_3 \}=\{R({\bf \Theta})\hat {\bf x},R({\bf \Theta})\hat {\bf y},R({\bf \Theta}) \hat {\bf z}\}$. Here
\begin{equation}
R_y(\alpha) = \left(
\begin{array}{c c c}
\cos \alpha & 0 & - \sin \alpha \\
0 & 1 & 0 \\
\sin \alpha & 0 & \cos \alpha\\
\end{array}
\right)
\quad \textrm{and} \quad
R_z(\alpha) = \left(
\begin{array}{c c c}
\cos \alpha  & \sin \alpha & 0 \\
- \sin \alpha & \cos \alpha & 0 \\
0 & 0 & 1\\
\end{array}
\right)
\end{equation}
matrices correspond to rotations around the fixed laboratory axes $\hat {\bf y}$ and $\hat {\bf z}$.

Because of the rotational and translational invariance, the free energy cost of ribbon deformations only depends on derivatives of the attached frame, which can be expressed as the rate of rotation ${\bf \Omega}(s)$ of the ribbon frame~\cite{moroz98, panyukov00}
\begin{equation}
\frac{d{\bf e}_i}{ds} = \frac{dR}{ds} R^{-1}  {\bf e}_i \equiv {\bf \Omega} \times {\bf e}_i, \quad \quad i=1,2,3.
\end{equation}
Here, deviation from flatness is measured by the components of ${\bf \Omega}=\Omega_i {\bf e}_i$, where $\Omega_1(s)$ and $\Omega_2(s)$ are the ribbon bending curvatures around axes ${\bf e}_1(s)$ and ${\bf e}_2(s)$, and $\Omega_3(s)$ is a twisting strain of the ribbon around the ${\bf e}_3(s)$ axis. Alternatively we can view $\Omega_i(s)$ as the rates of rotation of the ribbon about the axis ${\bf e}_i$ as a function of the arclength $s$. In terms of the Euler angles, the rates of rotation are
\begin{eqnarray}
\Omega_1 & = & \sin \psi \frac{d\theta}{ds} - \cos \psi \sin \theta \frac{d\phi}{ds},\nonumber \\
\Omega_2 & = & -\cos \psi \frac{d\theta}{ds} -\sin \psi \sin \theta \frac{d\phi}{ds},\nonumber \\
\Omega_3 & = & -\frac{d\psi}{ds} - \cos \theta \frac{d\phi}{ds}.
\end{eqnarray}
To the lowest order in ${\bf \Omega}(s)$, the energy cost of ribbon deformations can be expressed as~\cite{landauB}
\begin{equation}
E=\int\! \frac{ds}{2} \left[A_1 \Omega_1^2 + A_2 \Omega_2^2 + C \Omega_3^2 \right].
\end{equation}
If ribbon is constructed from a 3-dimensional isotropic elastic material of Young's modulus $E$ and Poisson's ration $\nu$, then the parameters $A_i$ are~\cite{landauB}
\begin{equation}
A_1 = E Wh^3/12, \quad A_2 = E W^3 h/12, \quad C = \mu_3 W h^3/3,
\end{equation}
where $\mu_3=E/2(1+\nu)$ is the 3-dimensional shear modulus and $\nu$ is the Poisson's ratio. In terms of the two-dimensional graphene elastic parameters $\kappa$, $Y$ and $\nu$ in the main text, we have
\begin{equation}
A_1 = \kappa W (1 - \nu^2), \quad A_2 = Y W^3/12, \quad C = 2 \kappa W (1-\nu).
\end{equation}
In the limit of large F\"oppl-von Karman number $Y W^2/\kappa \gg 1$, we find that $A_2 \gg A_1, C$. As was shown in the main text, for ribbons whose width $W$ is larger than the thermal length scale $\ell_\textrm{th} \sim \kappa/\sqrt{k_B T Y}$ the internal thermal fluctuations of the ribbon renormalize bending and twisting rigidities to
\begin{equation}
A_1 \sim \frac{\kappa W^{1+\eta}}{\ell_\textrm{th}^\eta}, \quad A_2 \sim Y W^{3-\eta_u}\ell_\textrm{th}^{\eta_u}, \quad C \sim \frac{\kappa W^{1+\eta}}{\ell_\textrm{th}^\eta},
\end{equation}
where $\eta\approx 0.82$ and $\eta_u = 2- 2 \eta \approx 0.36$. As noted there for graphene membranes the thermal length at room temperature is of order the lattice constant $\ell_\textrm{th} \sim 1 \textrm{\AA}$, so these renormalizations can be extremely large.

In the presence of an external edge force $F$ along the laboratory $z$-axis, the total free energy becomes 
\begin{equation}
E=\int\! \frac{ds}{2} \left[A_1 \Omega_1^2 + A_2 \Omega_2^2 + C \Omega_3^3 \right] -  F z,
\end{equation}
where $z=\int\! ds\,({\bf e}_3 \cdot \hat {\bf z})$ is the ribbon end-to-end separation in the $\hat {\bf z}$ direction. In the presence of thermal fluctuations, the expected value of $z$ is
\begin{equation}
\langle z \rangle = k_B T \frac{\partial \ln Z}{\partial F},
\label{eq:si:response_z_to_F}
\end{equation}
where we introduced the partition function 
\begin{equation}
Z=\int \mathcal{D}[{\bf \Theta}(s)] e^{-E/k_B T}.
\end{equation}

\subsection{Schr\"odinger like equation}
By the usual transfer matrix/path integral arguments for statistical mechanics in one dimension, the partition function $Z$ is closely related to the propagator for the probability distribution of ribbon frame orientation, where the unnormalized propagator is defined as
\begin{equation}
G({\bf \Theta}_f,s_f|{\bf \Theta}_i)=\int_{{\bf \Theta}(0)={\bf \Theta}_i}^{{\bf \Theta}(s_f)={\bf \Theta}_f} \mathcal{D}[{\bf \Theta}(s)] e^{-E/k_B T}.
\end{equation}
The function above propagates the initial distribution of Euler angles $\rho({\bf \Theta}_0,0)$ to 
\begin{equation}
\rho({\bf \Theta},s)=\int G({\bf \Theta},s|{\bf \Theta}_0) \rho({\bf \Theta}_0,0) d{\bf \Theta_0},
\end{equation}
where $\rho({\bf \Theta},s)$ is unnormalized and the partition function is expressed as
\begin{equation}
Z=\int d{\bf \Theta} \rho({\bf \Theta},L),
\end{equation}
with the Euler-angle measure $\int \!d{\bf \Theta} \equiv \int_0^{2 \pi} d\phi \int_0^\pi \sin \theta d\theta \int_0^{2 \pi} d\psi$.
In order to derive a differential equation for the propagator, we consider its evolution over a short ribbon segment $\delta s$:
\begin{equation}
G({\bf \Theta}_f,s_f + \delta s|{\bf \Theta}_i) = \int \! d{\bf \Theta}\  e^{-\delta E/k_B T} G({\bf \Theta},s_f|{\bf \Theta}_i)
\end{equation}
From the equation above we follow Ref.~\cite{moroz98} to derive an imaginary time Schr\"odinger equation for the propagator
\begin{equation}
\left(\frac{\partial}{\partial s} + H\right)  G= \delta(s) \delta({\bf \Theta}-{\bf \Theta}_0).
\end{equation}
where $H$ is the Hamiltonian defined as
\begin{equation}
H=\frac{k_B T}{2} \left(\frac{\hat J_1^2}{A_1} + \frac{\hat J_2^2}{A_2}+\frac{\hat J_3^2}{A_3} \right) - \frac{F ({\bf e}_3 \cdot \hat {\bf z})}{k_B T}.
\label{eq:hamiltonian}
\end{equation}
Here the $\{\hat J_i\}$ are the angular momentum operators around the ribbon frame axes ${\bf e}_i$, which can be expressed in terms of derivatives with respect to Euler angles~\cite{eslami08, landauBquantummechanics}.
The distribution of ribbon frame orientations obeys a similar differential equation
\begin{equation}
\left(\frac{\partial}{\partial s} + H\right) \rho = 0,\textrm{\quad for }s>0.
\label{eq:schrodinger}
\end{equation}
By expanding the distribution of initial ribbon frame orientation in eigen-distributions $\rho_a({\bf \Theta})$, where $H \rho_a = \lambda_a \rho_a$, the ribbon frame orientation distribution and the partition function can be expressed as
\begin{eqnarray}
\rho({\bf \Theta},s) &=& \sum_a \alpha_a e^{-\lambda_a s} \rho_a({\bf \Theta}), \nonumber \\
Z & = & \sum_a \alpha_a e^{-\lambda_a L} \int\! d{\bf \Theta} \rho_a({\bf \Theta}).
\label{eq:si:part_function}
\end{eqnarray}
In the thermodynamic limit of very long ribbons ($L \rightarrow \infty$) the term with the smallest eigenvalue $\lambda_a$ dominates in the partition function and the expected value for the end-to-end separation of the ribbon in $\hat {\bf z}$ direction becomes
\begin{equation}
\left< \frac{z}{L} \right> = - k_B T  \frac{\partial}{\partial F} (\min_a \lambda_a)
\label{eq:z_def}
\end{equation}

\subsection{Analogy with the rotating top in quantum mechanics}
To proceed further (and to derive results valid for finite $L$ as well as $L \rightarrow \infty$), we note that the differential equation (\ref{eq:schrodinger}) looks like a quantum Schr\"odinger equation for a rotating top in gravitational field proportional to $F$, where the coordinate $s$ acts like imaginary time~\cite{yamakawa76,moroz98}. Hence, we can borrow methods from quantum mechanics to find eigen-distributions $\rho_a$  and eigenvalues $\lambda_a$. For a rotating top it is convenient to expand eigen-distributions in the basis of Wigner D functions $D^J_{MK}({\bf \Theta})$~\cite{landauBquantummechanics} with a well defined total angular momentum $\hat J^2 D^J_{MK}({\bf \Theta}) = J (J+1) D^J_{MK}({\bf \Theta})$ and angular momentum projections along the ribbon tangent $\hat J_3 D^J_{MK}({\bf \Theta}) = K D^J_{MK}({\bf \Theta})$ and the $z$ axis $\hat J_z D^J_{MK}({\bf \Theta}) = M D^J_{MK}({\bf \Theta})$, i.e.
\begin{equation}
\rho_a({\bf \Theta}) = \sum_{J=0}^\infty \sum_{K=-J}^J \sum_{M=-J}^J C^{J}_{a,K,M} D^J_{MK}({\bf \Theta}).
\end{equation}
In order to evaluate the partition function $Z$ in Eq.~(\ref{eq:si:part_function}), we need to evaluate integrals like
\begin{equation}
\int d{\bf \Theta} \rho_a({\bf \Theta}) = \sum_{J=0}^\infty \sum_{K=-J}^J \sum_{M=-J}^J C^{J}_{a,K,M} \int d{\bf \Theta}D^J_{MK}({\bf \Theta}) = 8 \pi^2 C^0_{a,0,0}.
\end{equation} 
Note that only those eigen-distributions  $\rho_a({\bf \Theta})$, which have non-zero component $C^0_{a,0,0}$, contribute to the partition function $Z$. Since the Hamiltonian $H$ in Eq.~(\ref{eq:hamiltonian}) does not mix Wigner D functions with different $M$ quantum numbers~\cite{landauBquantummechanics}, we can restrict the search for eigen-distributions $\rho_a({\bf \Theta})$ to the subspace with $M=0$, where Wigner D matrices can be expressed in terms of the spherical harmonics
\begin{equation}
D^J_{0K}(\psi,\theta,\phi)=\sqrt{\frac{4 \pi}{2 J + 1}} Y_J^{K*}(\theta,\phi),
\end{equation}
where $^*$ denotes the complex conjugate. 
In order to avoid additional normalization factors, it is convenient to expand eigen-distributions in the basis of spherical harmonics
\begin{equation}
\rho_a(\psi,\theta,\phi) = \sum_{J=0}^\infty \sum_{K=-J}^J C_{a,J}^K Y_J^K(\theta,\phi).
\end{equation}

Then the eigenvalues $\lambda_a$ and corresponding eigen-distributions $\rho_a(\psi,\theta,\phi)$ can be found from the matrix equation
\begin{equation}
\sum_{J,K} (\langle J',K' | H | J,K \rangle -\lambda \delta_{J,J'} \delta_{K,K'}) C^K_{J}=0,
\end{equation}
where 
\begin{eqnarray}
\langle J',K' | H | J,K \rangle &=&  \int_0^\pi \! \sin\theta d\theta \int_0^{2 \pi}\!\! d\phi \ Y_{J'}^{K'*}(\theta,\phi) \ H \ Y_J^K(\theta,\phi), \nonumber \\
\langle J',K' | H | J,K \rangle &=& \frac{k_B T}{2} \delta_{J,J'} \left[  \delta_{K,K'}  \frac{(J(J+1)-K^2)}{2}  \left(\frac{1}{A_1} + \frac{1}{A_2} \right) +   \delta_{K,K'} \frac{K^2}{C}\right. \nonumber \\
&& +  \delta_{K-2,K'}  \frac{\sqrt{(J+K)(J+K-1)(J-K+1)(J-K+2)}}{4} \left(\frac{1}{A_1} - \frac{1}{A_2} \right) \nonumber \\
&& \left. +  \delta_{K+2,K'}  \frac{\sqrt{(J+K')(J+K'-1)(J-K'+1)(J-K'+2)}}{4} \left(\frac{1}{A_1} - \frac{1}{A_2} \right) \right] \nonumber \\
&& -\frac{F}{k_B T} \frac{ \delta_{K,K'} }{\sqrt{(2J+1)(2J'+1)}}\left[\delta_{J-1,J'} \sqrt{J^2-K^2} + \delta_{J+1,J'} \sqrt{J'^2-K^2}\right].
\end{eqnarray}
We have solved the above matrix equation numerically to find the whole spectrum of eigenvalues $\lambda_a$ and eigen-distributions $\rho_a(\psi,\theta,\phi)$.

In order to evaluate the partition function $Z$, we need to expand the initial ribbon orientation in terms of the eigen-distributions
\begin{equation}
\hat P \rho(\psi,\theta,\phi,s=0) = \int_0^{2 \pi} \!\! d\psi \ \rho(\psi,\theta,\phi,s=0)=\sum_{a} \alpha_a \rho_a(\psi,\theta,\phi),
\end{equation}
where $\hat P$ denotes projection to the $M=0$ subspace and
\begin{eqnarray}
\alpha_a &=& \int_0^{2 \pi} \!\! d\psi \int_0^\pi \!\sin\theta d\theta \int_0^{2 \pi} \!\!d\phi \ \rho^*_a(\psi,\theta,\phi) \rho(\psi,\theta,\phi,s=0) =
\sum_{J=0}^\infty \sum_{K=-J}^J C_{a,J}^{K*} c_J^{K} , \nonumber\\
c_J^K &=& \int_0^{2 \pi} \!\! d\psi  \int_0^\pi \!\sin\theta d\theta \int_0^{2 \pi} \!\!d\phi\  Y_J^{K*}(\theta,\phi)\  \rho(\psi,\theta,\phi,s=0) .
\end{eqnarray}
The partition function is then
\begin{eqnarray}
Z & = & \sum_a \alpha_a e^{-\lambda_a L} \int_0^{2 \pi} \!\! d\psi \int_0^\pi \!\sin\theta d\theta \int_0^{2 \pi} \!\!d\phi \ \rho_a(\psi,\theta,\phi), \nonumber \\
Z&=& \sum_a \alpha_a e^{-\lambda_a L} 4 \pi^{3/2} C_{a,0}^0.
\end{eqnarray}
Finally, the average ribbon end-to-end distance $\langle z \rangle$ is obtained by taking derivative of this partition function $Z$ with respect to force, see Eq.~(\ref{eq:si:response_z_to_F}).

As was mentioned in the main text, this same formalism can be used to study both the pulling and bending of ribbons. For pulling we orient the ribbon along the $\hat {\bf z}$ axis with the initial ribbon orientation ${\bf \Theta}_i = \{0,0,0\}$, which results in
\begin{equation}
c_J^K = Y_J^{K*}(\theta=0,\phi=0)= \delta_{K,0} \sqrt{\frac{2 J+1}{4 \pi}}.
\end{equation}
For bending around axis $\hat {\bf e}_1$ we orient the ribbon along the $\hat {\bf x}$ axis with the initial ribbon orientation ${\bf \Theta}_i = \{\pi/2,\pi/2,0\}$, which results in
\begin{equation}
c_J^K= Y_J^{K*}\left(\theta=\frac{\pi}{2},\phi=\frac{\pi}{2}\right) = (-1)^{(K+|K|)/2} \  \frac{2^{|K|} (-i)^K}{2\pi} \cos\left[\frac{\pi (J + |K|)}{2} \right] \sqrt{\frac{(2 J+1) (J-|K|)!}{(J+|K|)!}}\  \frac{\Gamma[(J+|K|+1)/2]}{\Gamma[(J-|K|+2)/2]}.
\end{equation}
For bending around axis $\hat {\bf e}_2$, which is harder because it involves the ribbon stretching, we orient the ribbon with the initial orientation ${\bf \Theta}_i = \{0,\pi/2,0\}$, which results in
\begin{equation}
c_J^K= Y_J^{K*}\left(\theta=\frac{\pi}{2},\phi=0\right) = (-1)^{(K+|K|)/2} \  \frac{2^{|K|}}{2\pi} \cos\left[\frac{\pi (J + |K|)}{2} \right] \sqrt{\frac{(2 J+1) (J-|K|)!}{(J+|K|)!}}\  \frac{\Gamma[(J+|K|+1)/2]}{\Gamma[(J-|K|+2)/2]}.
\end{equation}

\subsection{Large force limit}
For large pulling forces, we have to take into account both the ribbon stretching and the deformation energies that appear in
\begin{equation}
E=\int\! \frac{ds}{2} \left[A_1 \Omega_1^2 + A_2 \Omega_2^2 + C \Omega_3^3 + Y_{1D} u_{zz}^2 \right] -  \int\! ds \ F \left( \hat {\bf z} \cdot {\bf e}_3 \right) \left[1+ u_{zz} \right],
\end{equation}
where $u_{zz}$ corresponds to the stretching strain along the ribbon backbone, and the one dimensional Young's modulus is $Y_{1D}=Y_R(W) W$. Here, $Y_R(W)$ is the renormalized 2-dimensional Young's modulus evaluated at the scale of the ribbon width. For large pulling  forces the ribbon is nearly straight and the tangent ${\bf e}_3$ can be approximated as
\begin{equation}
{\bf e}_3 = t_x  \hat {\bf x} + t_y \hat{\bf y} + \left[1 - \frac{(t_x^2 + t_y^2)}{2}\right] \hat {\bf z},
\end{equation}
where $t_x, t_y \ll 1$. To quadratic order in $t_x$ and $t_y$, the free energy becomes
\begin{equation}
E = \int \! ds \left[\frac{A_1}{2} \left(\frac{\partial t_y}{\partial s} + t_x \Omega_3 \right)^2 + \frac{A_2}{2} \left(\frac{\partial t_x}{\partial s} - t_y \Omega_3 \right)^2 + \frac{C}{2} \Omega_3^2 + \frac{Y_{1D}}{2} u_{zz}^2 + \frac{F}{2} \left(t_x^2 + t_y^2 \right) - F u_{zz}\right].
\end{equation}
After integrating out the $\Omega_3$ and $u_{zz}$ the effective free energy becomes 
\begin{equation}
E_\textrm{eff} = \int \! \frac{ds}{2} \left[A_1 \left( \frac{\partial t_y}{\partial s} \right)^2 + A_2 \left( \frac{\partial t_x}{\partial s} \right)^2 + F\left(t_x^2 + t_y^2\right) - \frac{[A_1 t_x (\partial t_y/\partial s) - A_2 t_y (\partial t_x/\partial s) ]^2 }{(C + A_1 t_x^2 + A_2 t_y^2)}  \right].
\end{equation}
For $C>0$ the last term is 4-th order in $t_x$ and $t_y$ and can thus be neglected for large forces. Upon rewriting the effective free energy in Fourier space
\begin{equation}
E_\textrm{eff} = \frac{L}{2} \sum_{q} \left[ (F+ A_2 q^2 ) |t_x(q)|^2 + (F+ A_1 q^2) |t_y(q)|^2 \right],
\end{equation}
we find
\begin{equation}
\langle |t_x(q)|^2 \rangle = \frac{k_B T}{L (F+A_2 q^2)}, \quad \langle |t_y(q)|^2 \rangle = \frac{k_B T}{L (F+A_1 q^2)}.
\end{equation}
Using the results above we can find the ribbon extension
\begin{eqnarray}
\left<z \right> &=& \left< \int\! ds\left( \hat {\bf z} \cdot {\bf e}_3 \right) \left[1+ u_{zz} \right] \right> \approx
\left< \int \!ds\ \left[1 + u_{zz}- \frac{(t_x^2 + t_y^2)}{2}\right] \right >, \nonumber \\
\left<\frac{z}{L} \right> &=& 1 + \left< u_{zz} \right> - \frac{1}{2} \sum_q \left(\langle |t_x(q)|^2 \rangle + \langle |t_y(q)|^2 \rangle  \right), \nonumber \\
\left<\frac{z}{L} \right> & = & 1 + \frac{F}{Y_{1D}} - \frac{k_B T}{4 \sqrt{F}} \left(\frac{1}{\sqrt{A_1}} + \frac{1}{\sqrt{A_2}} \right).
\end{eqnarray}
We find that for large forces the ribbon extension is independent of the twisting rigidity $C$.

\end{widetext}

\end{document}